\def\ISLONG{0} 
\def\ISPREPRINT{1} 

\documentclass[sigconf]{aamas} 

\ifnum\ISPREPRINT=1  
    \renewcommand\footnotetextcopyrightpermission[1]{} 
\else{}
\fi

\usepackage{balance} 
\usepackage{algorithm}
\usepackage{algorithmic}
\usepackage{diagbox} 
\usepackage{multirow}
\usepackage{subfigure}

\setcopyright{ifaamas}
\acmConference[AAMAS '23]{Proc.\@ of the 22nd International Conference
on Autonomous Agents and Multiagent Systems (AAMAS 2023)}{May 29 -- June 2, 2023}
{London, United Kingdom}{A.~Ricci, W.~Yeoh, N.~Agmon, B.~An (eds.)}
\copyrightyear{2023}
\acmYear{2023}
\acmDOI{}
\acmPrice{}
\acmISBN{}

\title[Limit Order Book Simulation]{Neural Stochastic Agent-Based Limit Order Book Simulation: A Hybrid Methodology}

    \author{Zijian Shi}
    \affiliation{
      \institution{University of Bristol}
      \city{Bristol}
      \country{United Kingdom}}
    \email{zijian.shi@bristol.ac.uk}

    \author{John Cartlidge}
    \affiliation{
      \institution{University of Bristol}
      \city{Bristol}
      \country{United Kingdom}}
    \email{john.cartlidge@bristol.ac.uk}

\begin{abstract}
Modern financial exchanges use an electronic limit order book (LOB) to store bid and ask orders for a specific financial asset. As the most fine-grained information depicting the demand and supply of an asset, LOB data is essential in understanding market dynamics. Therefore, realistic LOB simulations offer a valuable methodology for explaining empirical properties of markets. Mainstream simulation models include agent-based models (ABMs) and stochastic models (SMs). However, ABMs tend not to be grounded on real historical data, while SMs tend not to enable dynamic agent-interaction. To overcome these limitations, we propose a novel hybrid LOB simulation paradigm characterised by: (1) representing the aggregation of market events' logic by a neural stochastic background trader that is pre-trained on historical LOB data through a neural point process model; and (2) embedding the background trader in a multi-agent simulation with other trading agents. We instantiate this hybrid NS-ABM model using the ABIDES platform. We first run the background trader in isolation and show that the simulated LOB can recreate a comprehensive list of stylised facts that demonstrate realistic market behaviour. We then introduce a population of `trend' and `value' trading agents, which interact with the background trader. We show that the stylised facts remain and we demonstrate order flow impact and financial herding behaviours that are in accordance with empirical observations of real markets. 
\end{abstract}



\newcommand{\BibTeX}{\rm B\kern-.05em{\sc i\kern-.025em b}\kern-.08em\TeX}

\begin{document}


\pagestyle{fancy}
\fancyhead{}


\maketitle 


\section{Introduction}

Modern financial exchanges (e.g., London Stock Exchange) and online commodity auction platforms (e.g., StockX) use the continuous double auction (CDA) mechanism to determine the price of assets. Buyers and sellers continuously submit orders to a limit order book (LOB), where order matching takes place and transactions result. 
Formally, a LOB is a continuously updating queueing system for limit orders. Bid orders (i.e., orders to buy) and ask orders (i.e., orders to sell) are queued on two sides of the book by price-time priority. Whenever order prices cross (i.e., price of the new bid order is higher than the lowest ask price, or price of the new ask order is lower than the highest bid price), orders are matched into transactions. 
Whenever a new order is submitted, or an existing order is cancelled the LOB will update. These updates happen continuously and time intervals between events can be in the order of nanoseconds. 
In this sense, the LOB depicts the most fundamental and fine-grained level of demand and supply information concerning a specific financial asset. 
Therefore, the LOB is often used as a primary data source in financial microstructure studies to investigate topics such as the price formation mechanism \cite{naes2006order}, stochastic properties of the market \cite{bouchaud2002statistical}, and the influence of high-frequency trading \cite{nolte2016high}.

However, although the LOB is of critical importance in research, using historical LOB data suffers from two major problems. 
First, as one cannot interact with historical data, it cannot be used to conduct `what if' counterfactual analysis. Therefore, backtesting a trading algorithm on historical data suffers from the unrealistic assumption that the market will not react. This is a particularly dangerous assumption when trading large volumes that will `move' the market. 
Second, the availability of historical LOB data is limited and expensive.\footnote{http://www.nasdaqtrader.com/Trader.aspx?id=DPUSdata} 
Simulations and synthetically generated data, respectively, can help to overcome these problems. 

Two mainstream models for LOB simulation and synthetic data generation are agent-based models (ABMs) and stochastic models (SMs). 
ABMs construct an interactive trading environment containing heterogeneous agents with behavioural logic that is either human-defined or learned. 
While defined agents are often modelled as simplifications of real-world actors such as market makers, momentum traders, and liquidity providers, the objectivity of ABMs can be doubted as one cannot guarantee that the simulation will behave in the same manner as the real world \cite{gould2013limit}.
On the other hand, SMs learn stochastic properties directly from historical data and so are objectively grounded in reality. Prior studies have shown that market characteristics, such as price and volume distributions and the arrival of market events, can be captured using SMs, and this learned knowledge can be used to generate realistic synthetic data. 
However, SMs have the limitation of not being interactive. Therefore, while SMs can generate endless amounts of synthetic data, the problems associated with backtesting on real data remain. 

This paper presents a neural stochastic agent-based model (NS-ABM) to simulate the LOB. NS-ABM is a novel hybrid methodology that combines the two general approaches of ABM and SM. Our primary contributions are summarised as:

\begin{enumerate}
    \item We introduce a neural stochastic background trader (BT) whose actions mimic the aggregation of order events posed by the whole market. The BT utilises an advanced state-dependent parallel neural Hawkes process model to learn the behavioural pattern of the whole market from historical real-world level-2 LOB data. When coupled with several empirical observations concerning order prices and volumes, the BT agent is able to stochastically sample realistic order event streams.
    
    \item We incorporate the BT into the open-source ABIDES \cite{byrd2020abides}  simulation framework. The BT is shown to produce LOB dynamics that reproduce a comprehensive list of ten stylised facts about real world LOBs. Such high fidelity has not been shown in previous ABMs.
    
    \item We introduce a population of trading agents with various `trend' and `value' trading strategies and demonstrate that the BT reacts realistically to endogenous events caused by other trading agents. The resultant LOB dynamics continue to exhibit the stylised facts of real markets, which demonstrates the realistic behaviour of the ABM. We also demonstrate that the ABM exhibits order flow impact and financial herding behaviours that are similar to empirical observations of real markets. 
\end{enumerate}


\section{Background and Related Works}

\subsection{Limit Order Book}
A CDA market allows market participants to submit both buy orders and sell orders for a specific asset, with no restrictions on time intervals between events. An electronic LOB is used to record submitted, yet unexecuted, orders. It is coupled with an order matching engine to match orders into transactions. As the LOB provides the most detailed demand and supply information in the market, it is considered to be the ultimate microscopic level of description \cite{bouchaud2002statistical}.

A {\em limit} order event $E(t,s,a,v,p)$ specifies a time of event $t$, a side of event $s\in\{bid,ask\}$, an action $a\in\{submission,cancellation\}$, a volume $v$, and a price $p$. The LOB is updated whenever a new event arrives.
The LOB contains a bid list and an ask list, each sorted by price-time priority such that the bid at the front of the bid list (i.e., the {\em best bid}) has the highest price, $p^{b(1)}$, and the ask at the front of the ask list (i.e., the {\em best ask}) has the lowest price, $p^{a(1)}$. Best bid $p^{b(1)}$ and best ask $p^{a(1)}$ are termed {\em quote prices}, and are considered as the {\em top price} level. 
The difference between $p^{b(1)}$ and $p^{a(1)}$ is named the \textit{bid-ask spread} and the average of $p^{b(1)}$ and $p^{a(1)}$ is named the \textit{mid-price}. We refer to $p^{b(n)}$ and $p^{a(n)}$, where $n\in\{2,3,\ldots\}$, as {\em deep price} levels.
When new bid $b$ arrives with price $p^{b}$ it will execute against the best ask if $p^{b} \geq p^{a(1)}$, else $b$ will enter the bid-side of the LOB in {\em descending} price-ordered position. Likewise, when new ask $a$ arrives with price $p^{a}$ it will execute against the best bid if $p^{a} \leq p^{b(1)}$, else $a$ will enter the ask-side of the LOB in {\em ascending} price-ordered position. 

\subsection{LOB Simulation}
LOB simulation offers a method to generate synthetic LOB data and perform trading experiments. 
Simulation is particularly useful for testing trading algorithms and explaining some of the empirical observations of real markets, for instance factors that lead to extreme price events \cite{paddrik2012agent} and how latency arbitrage affects market efficiency \cite{wah2016latency,duffin18}. Therefore, LOB simulation can provide insights for both market investors who want to maximise the profitability of their strategies, and market regulators who want to find reasons for market anomalies and take preventative measures. 
While a variety of LOB simulation methods exist, here we categorise into two general categories, agent based models (ABM) and stochastic models (SM).  

ABMs are a common method for performing LOB simulation. These bottom-up simulations include a virtual LOB venue in which heterogeneous trading agents interact, and LOB dynamics emerge from this system of interaction events. 
Often, agent types in ABMs are generalised simplifications of real market entities. For instance, a market maker agent imitates the role of large security broker-dealers that provide liquidity on both sides of the book; while a strategic momentum or mean reversion trader imitates the common strategy of investors to follow price trends. 
By rooting agent behaviours in reality, the aggregate effect of agents' interactions can lead to insightful findings that are in accordance with empirical studies of real world markets. For example, 
\citeauthor{cont2007volatility} \cite{cont2007volatility} used an ABM to investigate factors that cause volatility clustering in asset returns. Agents were configured to make trading decisions based entirely on perceived market volatility. Results showed that there is a link between frequency of market activity and agent threshold behaviour; and volatility clustering might be caused by investor inertia. 
\citeauthor{wang2021spoofing} \cite{wang2021spoofing} set up an ABM with zero intelligence (ZI) agents and heuristic belief learning (HBL) agents to investigate the role of spoofing in price manipulation. Both ZI agents and HBL agents use noisy observations on an exogeneous fundamental value to make trading decisions, with HBL agents exploiting additional LOB information to determine when prices are under/overvalued. Results showed that the LOB-dependent decision making of HBL agents mean they are more easily misled by spoofing and therefore price is more vulnerable to manipulation. 
\citeauthor{mcgroarty2019high} \cite{mcgroarty2019high} developed a realistic ABM containing several agent types including market makers, liquidity consumers, strategic traders, and noise traders. Data generated by the simulation exhibits several stylised facts found in real markets, which demonstrates realistic simulation behaviour.
However, despite the reported successes of ABMs, concerns have been expressed over their subjectivity. In particular, there is no guarantee that the real market will act in a similar way to any particular ABM, as the behavioral patterns and parameters of agents are subjectively determined. Also, some have argued that it may not be possible to accurately model an individual's complex behaviour though simple trading rules  \cite{preis2007statistical}.

SMs, on the other hand, simulate the LOB through stochastic assumptions or observations on the aggregate order flow. Unlike ABM, SM is fundamentally grounded on real data. A variety of stochastic models have been used to simulate the LOB. 
\citeauthor{cont2010stochastic} \cite{cont2010stochastic} used an independent Poisson processes to model order arrivals and cancellations. The model was shown to be capable of fast inference from historical data and enabled efficient calculation of conditional probabilities of events such as price movement. The simulated data generated was also shown to replicate several dynamic properties of the real LOB, despite having a minimal set of assumptions. 
In \cite{cont2013price}, the LOB was modelled as a Markovian queuing system. This form of model can provide analytical expressions for various quantities of interest, and has provided insights into the relation between price dynamics and order flow. 
Some research, on the other hand, took a deep learning perspective in stochastic LOB simulation. For instance, the LOB recreation model presented in \cite{shi2021limit,shi2021lobrm} modelled several stochastic properties of the LOB using a continuous variant of RNN to predict LOB volumes from top level trades and quotes data. The model was able to achieve a static simulation of the LOB by concatenating consecutive predictions while ignored the dynamic event-based characteristic of the system, leading to its failure to replicate market stylized facts.
More recently, \citeauthor{shi2022state} \cite{shi2022state} modeled the LOB event stream using a state-dependent parallel neural Hawkes process (sd-PNHP). The model was shown to exhibit superior performance over pure stochastic models when predicting the type and the time of next LOB event. Furthermore, the synthetic LOB data generated by the model was shown to be more realistic than previous models, as demonstrated by the number of stylised facts exhibited. Nevertheless, despite these successes, stochastic models have the shortcoming of not allowing dynamic interaction. 

In this paper, we introduce a hybrid SM-ABM. We hypothesise that this approach will offer the advantages of both models: an interactive, bottom-up LOB simulation that is grounded in real-world data and exhibits real-world characteristics.  
Others have previously suggested the advantages of combining models \cite{feng2012linking}, however, there are few prior studies that have attempted to embed stochastic models of the LOB in an interactive ABM environment.
These include  \citeauthor{panayi2015stochastic} \cite{panayi2015stochastic}, who claim their model is a hybrid of ABM and SM; however, the model is essentially a stochastic model in which order flows are attributed to imaginary agents, and there is no provision to interact with the stochastic system. 
More recently, \citeauthor{kumar2021deep} \cite{kumar2021deep} proposed an ABM model in which a stochastic market maker is embedded and the market maker interacts with other trading agents. However, the focus of that study is on deploying the stochastic model as a strategic trader to make profit; the model is not designed as a realistic market simulator to investigate market dynamics.
The most similar research that we have found is \cite{coletta2022learning}; in which a conditional generative adversarial network (CGAN) based agent is first trained on real data, and is then used as a `world agent'. The world agent is then incorporated into an ABM simulation to conduct interaction experiments. Nevertheless, CGAN agent lacks support from a statistical perspective, being a complete deep learning `black box'. Besides, the CGAN model cannot be validated through comparing prediction accuracy with mainstream models, while the sd-PNHP model can be rigorously grounded on real data using criteria like log likelihood of event arrival time and event type prediction accuracy.
Thus, we propose that there is a research gap, which we attempt to address in this paper. 


\section{Model Formulation}
The NS-ABM for LOB simulation has two main aspects. First, the simulation is agent-based and the ABM framework we use is the ABIDES open-source LOB simulation. Second, we train a neural stochastic `background trader' whose behaviour logic is learned through a sd-PNHP model on historical trading data. We then incorporate the BT as an agent into ABIDES, and introduce other agent traders that can interact with the BT. 

\subsection{Agent-Based Interactive Discrete Event Simulation (ABIDES)}
\begin{figure}
    \centering
    \includegraphics[width=\linewidth]{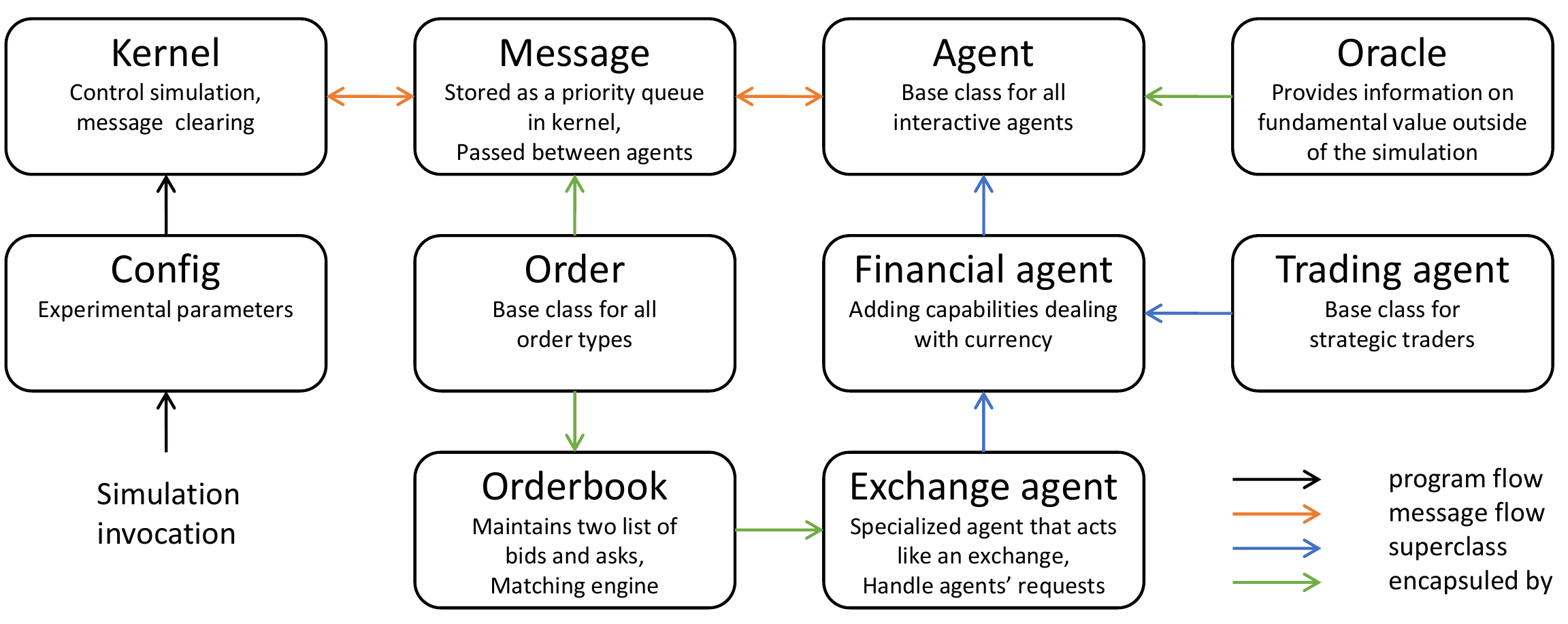}
    \caption{Simulation workflow of ABIDES; adapted from \cite{byrd2020abides}.}
    \label{fig:workflow}
\end{figure}

ABIDES \cite{byrd2020abides} is an agent-based simulation framework used to generate high-fidelity LOB data and conduct microstructure experiments. It has been adopted in varies studies, e.g., for developing and evaluating trading agents \cite{karpe2020multi}, and for investigating market manipulation \cite{yagemann2020feasibility}. ABIDES mimics real market settings in several ways: (1) it has a realistic messaging system derived from NASDAQ's published equity trading protocols, ITCH and OUCH; (2) it has no assumptions or restrictions on market settings, such as order size or time intervals between discrete events; and (3) it is equipped with a set of classes and functions that enable extension of existing agent types and actions. Fig.~\ref{fig:workflow} is a simplified version of the simulation logic shown in \cite{byrd2020abides}. In the following paragraphs, we introduce two classes of ABIDES' high frequency trading (HFT) agents that we use in this paper: `trend' trading agents and `value' trading agents.

Trend agents assume that a trend in price will persist or reverse in the near future, and they are called momentum (MM) agents and mean reversion (MR) agents, respectively. Trend agents calculate the current trend in price using the difference between moving average price over a long time window and over a short time window. The trading logic of trend agents is illustrated in Algorithm~\ref{alg:trend}. These agents try to maintain a neutral position, so the submitted order size is either equal to the size that can close the current position, otherwise a standard size $u$ is used. For instance, assume an agent has a short stock holding of $-200$. If the agent then decides to submit a market buy order, the order size would be $200$ to close the position; conversely, if the agent decides to submit a market sell order, the order size would be $u$ to further expand the position. By maintaining a neutral position, trend agents attempt to profit from high-frequency trading; rather than profit from capital gains by holding a large position over a comparatively long time horizon.  

\begin{algorithm}[t]
\caption{Trading logic for trend agents (MM \& MR)}
\small
\label{alg:trend}
\begin{algorithmic}[1] 
\REQUIRE window size $l_{1}$ and $l_{2}$, $l_{2}>l_{1}$, current stock holding $h$
\IF {receive wake up call from the kernel}
    \STATE cancel all unexecuted orders    
    \STATE send request for quote prices to the exchange agent
    \STATE receive quote prices $(p_{t}^{a(1)},p_{t}^{b(1)})$ from the exchange agent
    \STATE calculate mid price $p_{t}^{mid}$ and append it to the agent's internally stored list $mid\_list$
    \IF{$\mathrm{len}(mid\_list)>=l_{2}$}
        \IF{$\mathrm{MA}_{l_{1}}(mid\_list)>\mathrm{MA}_{l_{2}}(mid\_list)$} 
            \IF{$\mathrm{type}(\mathrm{agent})==MM$}
                \STATE submit a market buy order, $q=\mathrm{max}(-h,u)$
            \ELSIF{$\mathrm{type}(\mathrm{agent})==MR$}
                \STATE submit a market sell order, $q=\mathrm{max}(h,u)$
            \ENDIF
        \ELSE 
            \STATE do steps 8-11 using inverse logic (i.e., MM sells, MR buys)
        \ENDIF
    \ELSE 
        \STATE queue next wake up call in kernel, break
    \ENDIF
    \STATE queue next wake up call in kernel 
\ENDIF
\end{algorithmic}
\end{algorithm}

Value agents trade according to observations on a fundamental value oracle that is exogeneous to the market \cite{byrd2019explaining}. Such an oracle can be any time series generated through a stochastic process. We use the SparseMeanRevertingOracle class to generate fundamental values for ZI/HBL (detailed in \cite{byrd2019explaining}). As presented in \cite{byrd2019explaining}, the fundamental value follows an Ornstein-Uhlenbeck process, of which the value at $t_{n}$ is denoted as:

\begin{align}
    p_{t_{n}}=\mu +(p_{t_{n-1}}-\mu)e^{-\gamma \delta_{n}}+u_{t_{n}}\\
    u_{t_{n}}\sim N(0,\frac{\sigma^{2}}{2\gamma}(1-e^{-2\gamma \delta_{n}}))
\end{align}
in which $\delta_{n}$ denotes units of time elapsed since last observation. When $\delta_{n}$ increases, the fundamental value and its variance at $t_{n}$ converge to $\mu$ and $\frac{\sigma^{2}}{2\gamma}$, indicating its mean-reverting essence. 

Before agents observing the fundamental value, they first update internal estimates $\tilde{p}_{t_{n}}$ and $\tilde{\sigma_{t_{n}}}^{2}$:
\begin{align}
    \tilde{p}_{t_{n}}=(1-(1-\gamma)^{\delta_{n}})\mu+(1-\gamma)^{\delta_{n}}\tilde{p}_{t_{n-1}}\\
    \tilde{\sigma}_{t_{n}}^{2}=\frac{1-(1-\gamma)^{2\delta_{n}}}{1-(1-\gamma)^2}\sigma^{2}+(1-\gamma)^{2\delta_{n}}\tilde{\sigma}_{t_{n-1}}^{2}
\end{align}

After observing the fundamental value as $o_{t_{n}}$ with noise $\sigma_{o}^{2}$, estimates are updated in a Bayesian manner:
\begin{gather}
    \tilde{p}_{t_{n}} \leftarrow \frac{\sigma_{o}^{2}}{\sigma_{o}^2+\tilde{\sigma}_{t_{n}}^{2}}\tilde{p}_{t_{n}}+\frac{\tilde{\sigma}_{t_{n}}^{2}}{\sigma_{o}^2+\tilde{\sigma}_{t_{n}}^{2}}o_{t_{n}}\\
    \tilde{\sigma}_{t_{n}}^{2} \leftarrow \frac{\tilde{\sigma}_{t_{n}}^{2}\sigma_{o}^{2}}{\tilde{\sigma}_{t_{n}}^{2}+\sigma_{o}^{2}} 
\end{gather}
Upon finishing updating its interval values, the agent makes prediction on the fundamental value at $t_{n}+t_{w}$ as:

\begin{equation}
    \tilde{p}_{t_{n}+\Delta t}=(1-(1-\gamma)^{\Delta t})\mu+(1-\gamma)^{\Delta t}\tilde{p}_{t_{n}}
\end{equation}
of which is the value that value agents used to compare with the realized stock price at $t_{n}$ to make trading decision. Here the agents focus on the profit over short terms (in $\Delta t$), the same as momentum traders. The estimated fundamental value represents the `fair price' of the stock to the agent. If the current price on the LOB is underestimated, the agent tends to buy; otherwise the agent tends to sell. One type of value agent makes decisions based entirely on the observed fundamental value, and we call these zero intelligence (ZI) agents; the other type of value agent use additional information from the LOB to improve the possibility of order execution, and we call these heuristic belief learning (HBL) agents. We intentionally remove the private value setting over holding preference. As a result, the agents do not have preference over a long, to force agents to solely focus on profitability. That is, when there is no price information the agent has no preference over holding $100$ units or $-100$ units of stock. As trend agents do not have the private value setting in the ABDIES implementation, the removal of this setting also allows better comparison. The trading logic of value agents is described in Algorithm~\ref{alg:value}.

\begin{algorithm}[t]
\caption{Trading logic for value agents (ZI \& HBL)}
\small
\label{alg:value}
\begin{algorithmic}[1] 
\REQUIRE Intended maximum surplus $r_{max}$, look back period $l$, current stock holding $h$
\IF {receive wake up call from the kernel}
    \STATE cancel all unexecuted orders
    \STATE send request for quote prices to the exchange agent
    \STATE receive quote prices $(p_{t}^{a(1)},p_{t}^{b(1)})$ from the exchange agent
    \STATE obtain a noisy observation from the oracle, and update internal estimation on current value $p_{t}$ in a Bayesian manner
    \STATE make estimation on future fundamental value $p_{t+\Delta t}$
    \STATE sample requested surplus $r \sim \mathrm{Unif}(0,r_{max})$

    \IF{$\mathrm{rand}(0,1)>0.5$}
        \STATE submit a limit buy order, $q=\mathrm{max}(-h,u)$
            \IF{$\mathrm{type}(\mathrm{agent})==ZI$}
                \IF{$p_{t+\Delta t} - p_{t}^{a(1)}>r$}
                    \STATE $p=p_{t}^{a(1)}$  
                \ELSE 
                    \STATE $p=p_{t+\Delta t}-r$
                \ENDIF
            \ELSIF{$\mathrm{type}(\mathrm{agent})==HBL$}
                \STATE calculate execution probability vector $Prob$ at all price levels $P=(p_{min},...,p_{max})$ during past $L$ transactions
                \STATE calculate expected surplus $S=Prob*(p_{t+\Delta t}-P)$
                \STATE $p=p_{min}+\mathrm{argmax} S$
            \ENDIF 
    \ELSE 
        \STATE do steps 9-19 using inverse logic (i.e., submit a limit sell)
    \ENDIF
    \STATE queue next wake up call in kernel 
\ENDIF
\end{algorithmic}
\end{algorithm}

Trend agent and value agent types are representative of common trading strategies and have been widely studied in both analytical models and empirical studies. To aid analysis, we have deliberately chosen a minimal set of trading strategies, therefore these are the only two types of strategic trading agents that we consider in this work. However, it is trivial to include other pre-defined agent types, or define new trading strategies within ABIDES. We reserve such explorations of more complex markets for future work.

\subsection{Neural Stochastic Background Trader}

\begin{figure*}[tb]
\centering
\includegraphics[width=0.65\linewidth]{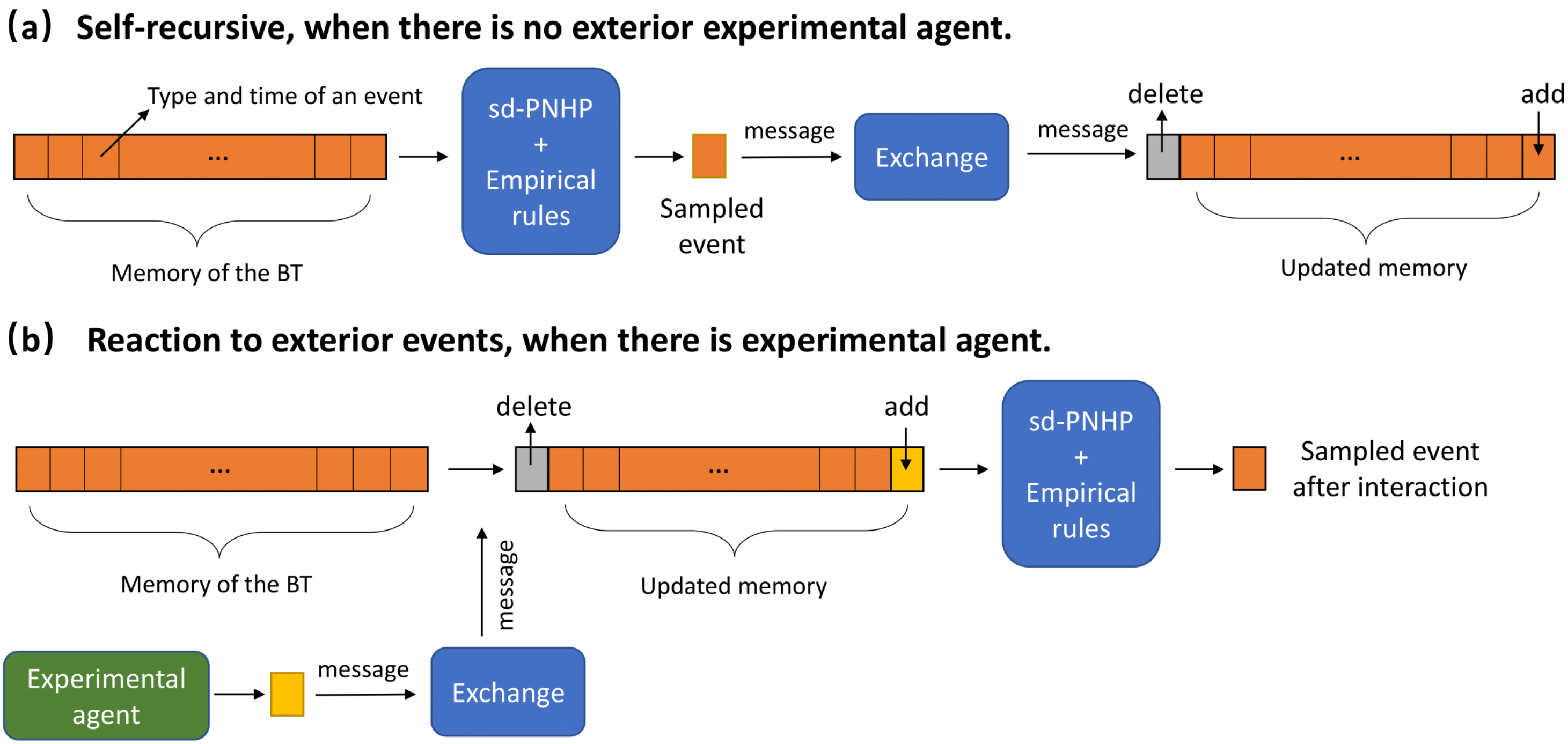}
\caption{(a) shows how the BT samples events when it is running on its own. A defined length of memory is used as input into the sd-PNHP model, coupled with empirical distributions concerning orders, to sample a new event. The event is then sent to the exchange, once accepted will be updated into its memory to go forward. (b) shows how interactions happen. It follows a similar logic of (a), the difference is that the BT will first combine experimental agent’s event into its memory, and then to sample events. The sampled event thus is influenced by other agents’ behaviours.} \label{fig:logic}
\end{figure*}

Pure ABMs in which all agents act according to human-defined rules can lack objective grounding. Although the simulation framework can be designed to closely imitate the real world (e.g., in ABIDES the messaging mechanism originates from NASDAQ, the LOB operates using real world rules, and the terminology of agent strategies can be found in real markets), realistic dynamics are not guaranteed. Critically, the behavioural patterns of agents, the parameters that determine those behaviours, and the complex interactions between agents combine to generate the overall order stream pattern. The parameters of individual agent behaviours cannot be learned from the aggregate order stream of real markets as the actions of individual traders are not known (data is anonymous and actions cannot be linked to individuals). However, as demonstrated by \cite{shi2022state}, it {\em is} possible to learn the overall order stream pattern of the whole market using a sd-PNHP model. Here, we adapt the approach of \cite{shi2022state} to learn aggregate order streams of real data and then deploy the model as an autonomous agent within an ABM so that it can interact and adapt to the actions of other trading agents in the market. We hypothesise that this `hybrid' ABM approach can generate more realistic market dynamics within which new trading strategies can be more rigorously tested.  

To this end, we implement a neural stochastic BT in ABIDES. The agent is backed with a sd-PNHP model that is trained on historical event stream data of the whole market. The agent can make predictions on next event type and next event time based on a defined length of event history. By iteratively incorporating newly sampled events into memory as input, the agent is able to endlessly generate event stream data that closely mimics real market data. We name the agent a `background trader' because: (1) when the agent is running on its own (as the only agent in the simulation), the order stream it generates represents the logic of a complete market; (2) when the agent interacts with other trading agents (i.e., when part of a multi-agent simulation), the reactions it produces are similar to aggregate market responses. Fig.~\ref{fig:logic} illustrates the logic of the BT when it is running on its own, and when it is interacting with other agents. By defining the BT as a subclass of the trading agent class in ABIDES, it inherits full functionality of communicating with the exchange, submitting and cancelling orders, and keeping track of its real-time profitability. We detail the logic of the neural stochastic trader in the following paragraphs. 

For a stochastic Hawkes process, the intensity rate of event arrival at a given time is based on a mean rate $\mu$, plus the additive exponentially decayable impact produced by historical events (controlled by $\alpha$ and $\delta$), as shown in Eq.~\ref{eq:1}. In contrast, in a multi-variate sd-PNHP model, the intensity rate for type $k$ event at a given time is decoded from the continuous latent state from the $k$-th continuous-time LSTM unit, with a defined length of past events and market states as input, as shown in Eq.~\ref{eq:2}.

\begin{equation} \label{eq:1}
    \lambda(t)=\mu+\sum_{h:t_{h}<t}\alpha*exp(-\delta(t-t_{h}))
\end{equation}

\begin{equation} \label{eq:2}
    \lambda_{k}(t)=\mathrm{D}_{k}(h_{k}(t))=\mathrm{D}_{k}(\mathrm{CTLSTM}_{k}(S,X,t))
\end{equation}

\begin{equation}\label{eq:3}
    p_{j}(t)=P(t_{j}=t \mid F^{S,X}_{j-})=\lambda(t) \mathrm{exp} \left (-\int_{t_{j-1}}^{t}\lambda(s)ds \right )
\end{equation}

According to fundamentals in the theory of stochastic process, the probability density function for event arrival time can be denoted as Eq.~\ref{eq:3}. After the model is well-trained on historical data, a  sampling method like the Ogata's thinning algorithm can be used to sample the most likely next event type and arrival time. Experiments performed in \cite{shi2022state} demonstrated that: (1) the model can make predictions with 50\%-60\% accuracy in event type prediction (compared with baseline accuracy of 25\% for this four class classification problem, with classes: ask submission, ask cancellation, bid submission, and bid cancellation); and (2) the model can be used to iteratively sample event streams. When combined with several empirical distributions of order price and order volume, it can be used to generate high-fidelity LOB data that exhibits multiple stylised facts found in real data.



\subsection{Implementation Details}
A first attempt of using the sd-PNHP as a pure SM to stochastically sample LOB event streams and conduct LOB simulation was illustrated in \cite{shi2022state}. Nevertheless, the original implementation was naive and lack essential settings to be interactive with experimental agents. 

First, previously submitted orders by the BT were not tracked by the LOB through indexing. Given an initial state of the LOB, the implementation was only adding or removing certain volumes of limit orders from the book whenever a new event was sampled, regardless of how the volumes on a level price were specifically composed of previously submitted orders. This would lead us to troubles when we need to identify which previously submitted orders are being transacted or cancelled, and to whom the transacted or cancelled orders belong. By incorporating the BT into ABIDES, both the BT and the exchange index the sampled orders. This allows the system to track a specific order from its submission to its cancellation or transaction, and also send according information to the originating agents using the agent-order mapping stored in the exchange. 

Second, as the sd-PNHP model focus on learning order patterns from a LOB of five price levels, default volumes (historical average value) are used when the LOB moves to a previously unseen price level. To ensure that all orders are generated from agents and are traceable by the exchange, this setting is removed. Recall that the BT is only able to generate and respond to events relating to the top five price level of the LOB. If the LOB has moved to a previously unseen price level (e.g. the 6-th best ask at time $t_{n-1}$ becomes the 5-th best ask at time $t_{n}$), the order volume on that price level starts from zero; If the LOB has left a price level behind (e.g. the 5-th best ask at time $t_{n-1}$ becomes the 6-th best ask at time $t_{n}$), orders on that price level are gradually removed from the LOB as cancellations which will not cause further response from the BT (as they are already out of the perception field of the BT).

Concerning more implementation of the BT in ABIDES, We use the base class of TradingAgent to build the BT. Main functionalities of the base class include being able to receive and send message (including receiving market information from the exchange, and posting order-related actions to the exchange), submit and cancel orders, and maintain its own order history. The BT is subscribed to market information. Whenever the LOB is updated resulting from its own or experimental agent’s action, the BT receives a market subscription message and samples an event (with event type, order statistics, and arrival time). A sampled event is not necessarily to be posed to the exchange, as there is possibility an event posed by other experimental agents can arrive earlier. Under this circumstance, the BT will sample a new event and the original event will be abandoned. The newly sampled event is deemed as an event under interaction (as in Fig.\ref{fig:logic}). There also exists emergency settings that in case the simulation encounters error. During market open, when there are no orders on the ask (or bid) side, the spread will be assumed to be one tick. If the BT samples an ask (or bid) cancellation action, it will be automatically replaced by an emergency refill action of submitting bid (or ask) limit order of volume 100 at the top price level. This mechanism is rarely activated, only when one side of the LOB runs out of liquidity.

Essential parameters regarding the BT are illustrated in Table \ref{tab:params} as ‘initial values’. Those values are either learned from data using MLE, or chosen  When market opens, the first five price levels are randomly populated with limit orders (volumes 100 - 1000) until the order volumes on each price levels reach an random aggregation value (volumes $1.5\times10^{4}$ - $2.0\times10^{4}$, indicated by empirical average value on top five price levels in the dataset). Time intervals for these orders are set as $10^{-6}$ second for fast population, during which time no market information will be sent out and no agents will react. These orders can be deemed as pre-market-open orders, and the resulted LOB is used as an initial state on which following events accumulate.

\section{Experiments}
\subsection{Model Learning and Parameter Setting}
The learning of the sd-PNHP model is based on the LOBSTER dataset, as in \cite{shi2022state}. The dataset contains real world LOB event stream data of five days' length for three stocks, ticker symbol INTC (Intel), MSFT (Microsoft), and JPM (JP Morgan), provided by the financial data provider LOBSTER.\footnote{A sample dataset can be found at \url{https://lobsterdata.com/info/DataSamples.php}} On average there are 0.5 million event updates per trading day per stock, and the LOB data is of five price levels, instead of being full market-depth. We choose the model trained on INTC data as the main model to be used in the following experiments.

Parameters in power law distributions for price and volume are learned on real data using maximum likelihood estimation. Other parameters, such as the proportion of all orders that are market orders and the probability that a limit order will shrink the bid-ask spread, are also estimated from real data. In a nutshell, for the BT, the sd-PNHP model decides the order stream pattern, and the learned stochastic parameters decides order-specific statistics. Parameters for experimental agents are human defined, taking reference from original settings in ABIDES.

\subsection{Configurations for ABIDES}

\subsection{Sensitivity Analysis}
First, we conduct sensitivity analysis on a system in which all order streams are generated by the BT. We consider key parameters that relate to the empirical distributions of order price and volume, the percentage of market orders in all orders, and some control parameters that enforce the book will not run out of liquidity. One thing to be noticed is that neural parameters that dominate the order stream pattern are not involved in sensitivity analysis, owing to its feature of being a `black box'. The exact parameters involved are listed in Table~\ref{tab:params}, and they are: (1) $P$: the exponent of power law distribution for order price; (2) $V1$: the exponent of power law distribution for order volume at the top price level (quote prices); (3) $V2$: the exponent of power law distribution for order volume at deep price levels (prices inferior to quote prices); (4) $Mi$: the market order imbalance index (0 indicating balanced, and $\pm 100\%$ indicating all market orders are bid or ask orders); (5) $Mv$: the exponent of power law distribution for market order volume; (6) $Lb$: the lower bound value for volumes at each price level. Once volumes fall below this value, upcoming limit order volumes will be forced to increase 1000 to ensure adequate liquidity in the market; (7) $Ip$: the possibility for a limit order to fall within the spread (being one tick higher than the best bid or one tick lower than the best ask) when the spread is larger than one. Spread is denoted by $sp$.

\begin{table}[t]
\caption{Parameters for sensitivity analysis}\label{tab:params}
\small
\begin{tabular}{lcc}
\toprule
\multicolumn{1}{c}{Parameter} & \begin{tabular}[c]{@{}c@{}}Initial value\\ ($sp=1$ / $sp>1$)\end{tabular} & \begin{tabular}[c]{@{}c@{}}Fluctuation\\range\end{tabular} \\ \midrule
$P$ - Price distribution   & 1.5 / 4.7  & $\pm$ 0.25   \\
$V1$ - Volume dist. (top) & 0.9 $\sim$1.2 & $\pm$ 0.25 \\
$V2$ - Volume dist. (deep) & 0 $\sim$1.8 & $\pm$ 0.25  \\
$Mi$ - Market imbalance & 0 & $\pm$ 100\%          \\
$Mv$ - Market vol. dist. & 1.2 / 1.6 & $\pm$ 0.25  \\
$Lb$ - Lower bound & 12500 & $\pm$ 2500            \\
$Ip$ - Inner spread prob. & 0.05  & $\pm$ 0.025    \\ 
\bottomrule
\end{tabular}
\end{table}

\begin{figure}[t]
    \centering
    \hspace{-10mm}
    \includegraphics[width=0.9\linewidth]{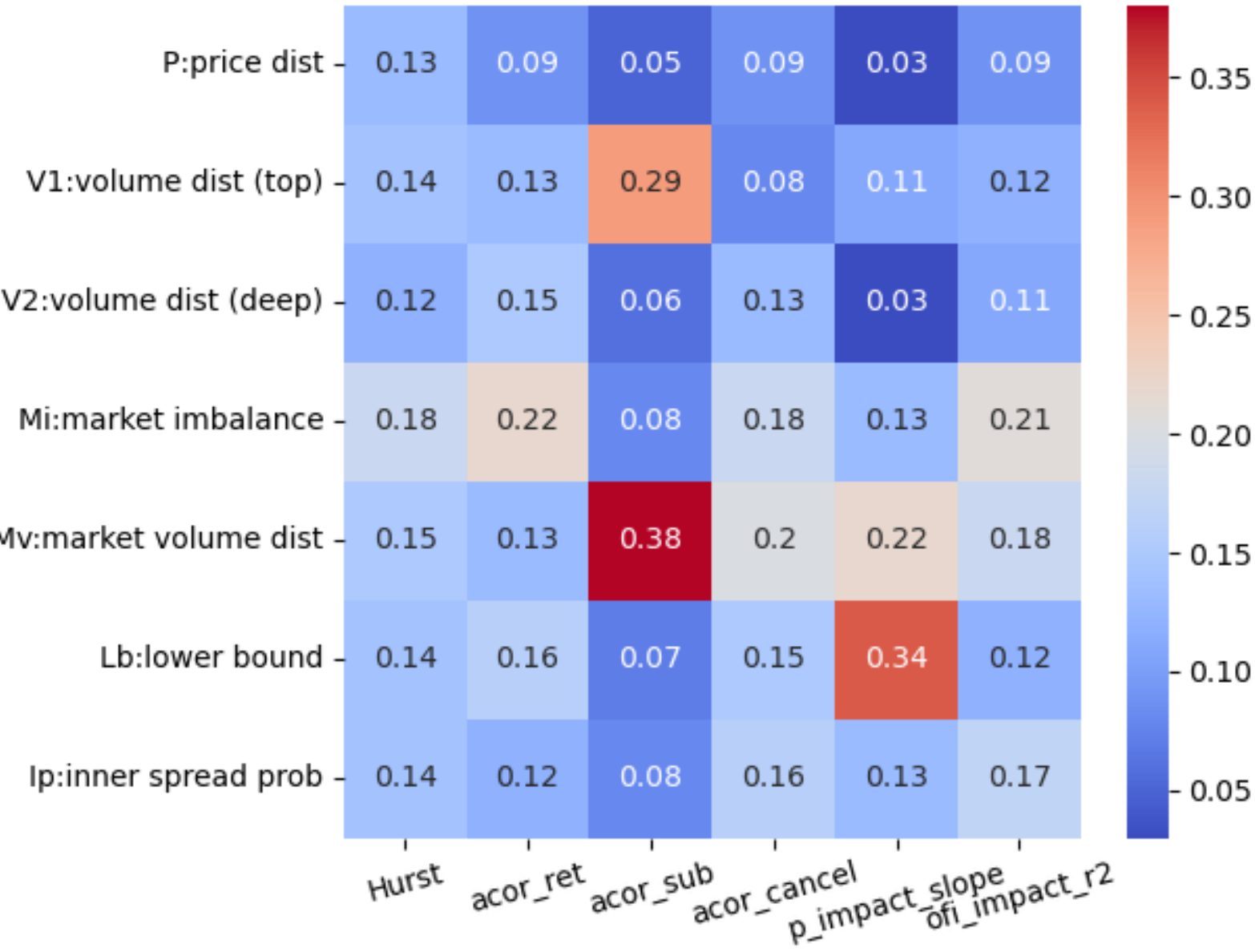}
    \caption{Heatmap for sensitivity analysis.}
    \label{fig:heatmap}
\end{figure}

We follow Sobol's global sensitivity analysis, as performed in \cite{mcgroarty2019high}. In this variance-based analysis, the ANOVA representation of a function $f(x)$ of $x=(x_{1},...,x_{n})$ is:
\begin{equation}
    f(x)=f_{0}+\sum_{s=1}^{n}\sum_{i_{1}<...<i_{s}}f_{i_{i}...i_{s}}(x_{i_{1}},...,x_{i_{s}})
\end{equation}
in which the $s$-th summation denotes the $s$-th order cooperative effect of variables $x_{i_{1}},...,x_{i_{s}}$ on the function output. Thus, the total variance and partial variance of $f(x)$ can be denoted as:

\begin{align}
    & D=\int(f(x)-f_{0})^2\rho (x)dx\\
    & D_{i_{1}...i_{s}}=\int f_{i_{1}...i_{s}}\rho(x_{i_{1}},...,x_{i_{s}})dx
\end{align}
The total variance concerning variable $x_{i}$ and its total sensitivity index can be calculated as:

\begin{align}
    & D_{i}^{tot} = \sum_{s=1}^{n} \sum_{\left \langle i \right \rangle}D_{i_{1}...i_{s}} \\
    & S_{i}^{tot}=\frac{D_{i}^{tot}}{D}
\end{align}
in which the symbol $\left \langle i \right \rangle$ sums over all $D$ terms that contain $i$. Sobol \cite{sobol2001global} provides an efficient Monte-Carlo method for calculating the indices. According to the method, $x$ is uniformly sampled from the input space. We then add random perturbations to each input element of $x$, and use all these $x$ to simulate $\mathrm{dim}(x)+1$ copies of LOBs. Each LOB is equivalent to one hour's length, consisting of roughly 0.3 million event updates. In total, $100*(\mathrm{dim}(x)+1)$ LOB samples are generated. Selected stochastic properties of the LOB are used to evaluate the sensitivity of the system, as in \cite{mcgroarty2019high}: (1) the Hurst exponent of volatility \cite{lillo2004long, gu2009emergence}; (2) the autocorrelation of mid-price return \cite{bouchaud2003theory}; (3) the first lag autocorrelation of order-sign for order submissions \cite{lillo2004long};  (4) the first lag autocorrelation of order-sign for order cancellations \cite{lillo2004long}; (5) the best exponent $\beta$ of the price impact function \cite{lillo2003master}; and (6) the R-squared for the order flow imbalance function \cite{cont2014price}. The first four criteria relate to the memory of order flow or the resulting mid-price series, and the last two criteria relate to the market price formation mechanism. Fig.~\ref{fig:heatmap} presents a heatmap after standardisation of the sensitivity indices.

From Fig.~\ref{fig:heatmap}, it can be seen that the top two parameters that have influence over the stochastic properties considered are: (1) the exponent of distribution for market order volume; and (2) market order percentage imbalance. In \cite{mcgroarty2019high}, it was argued that the upper limit of market order volume distribution is the most influential parameter, and the stochastic properties that it affects most are the Hurst exponent and the exponent for price impact function. Both the exponent for market order volume distribution and the upper limit for market order volume directly affect the size of market orders. Thus, we find that \cite{mcgroarty2019high}'s conclusions concur with our findings, even though the exact methodologies adopted differ.

\subsection{Stylised Facts}\label{sec:stylisedfacts}
`Stylised facts' in economics are empirical findings that are so consistent (for example, across a wide range of instruments, markets, and time periods) that they are accepted as truth. Such facts can be used to verify the fidelity of an economic simulation. In this section, we consider the behaviour of the simulated LOB against a comprehensive list of more than ten stylised facts. We generate all order streams by the BT in isolation, using the parameter settings shown in Table~\ref{tab:params} with fluctuation set to zero (i.e., all values are those initially learned; they are not perturbed). Results are averaged across 10 samples for both simulated and real data.

We were able to reproduce the following stylised facts:

\begin{description}
    
    \item[Hurst exponent for absolute return.] This fact indicates th\-at whether long-range memory exists in financial market time series \cite{cont2001empirical}.
    A detrended fluctuation analysis \cite{peng1994mosaic} can be applied on the time series of absolute returns to calculate the Hurst exponent. A Hurst exponent in the range of $(0.5,1)$ indicates the existence of long memory. Empirical studies indicate the Hurst exponent to be larger than 0.5 in stock markets \cite{gu2009emergence}. We find the exponent in simulated data to be 0.53, and 0.61 in the real data. 
    
    \item[Autocorrelation in order-sign series.] This fact indicates th\-at positive autocorrelation exists in order-sign series of submissions and cancellations, respectively. Empirical studies indicate the autocorrelation coefficient roughly falls in the range of $(0.2,0.3)$ \cite{lillo2004long}. We find the autocorrelation coefficients to be significant in both simulated and real data. In simulated data we find the coefficients to be 0.25 for submissions and 0.18 for cancellations. For real data, the coefficients are 0.41 and 0.35 respectively. 
    
    \item[Order flow imbalance impact.] This fact indicates that the order flow imbalance (OFI) tends to cause prices to change \cite{cont2014price}. The imbalance between supply and demand is measured as the difference between events that enforce the bid side (bid submission and ask cancellation), and events that enforce the ask side (ask submission and bid cancellation) during a ten-seconds interval. The R-squared value from the regression between return and OFI was found to have an average value of 0.65 in \cite{cont2014price}. Here, the R-squared value is found to be 0.64 in simulated data and 0.68 in real data.
    
    \item[Price impact function.] This fact indicates that the transaction volume's influence on price change is concave. Empirical studies indicate that the impact of transaction volume on change in quote prices increases more quickly with changes at small volumes and less quickly at larger volumes. The slope of the fitted curve between logarithm volume and logarithm price change ranges in $(0.1,0.5)$, and the slope varies across different markets owing to market protocols \cite{lillo2003master}. The slope calculated on simulated data is 0.25, and on real data is 0.11, both of which conform with a concave curve.
\end{description}

\begin{figure*}[t]
\centering
\subfigure[Mid-price evolution of simulation and real data.]{\label{fig:1}\includegraphics[width=0.4\linewidth]{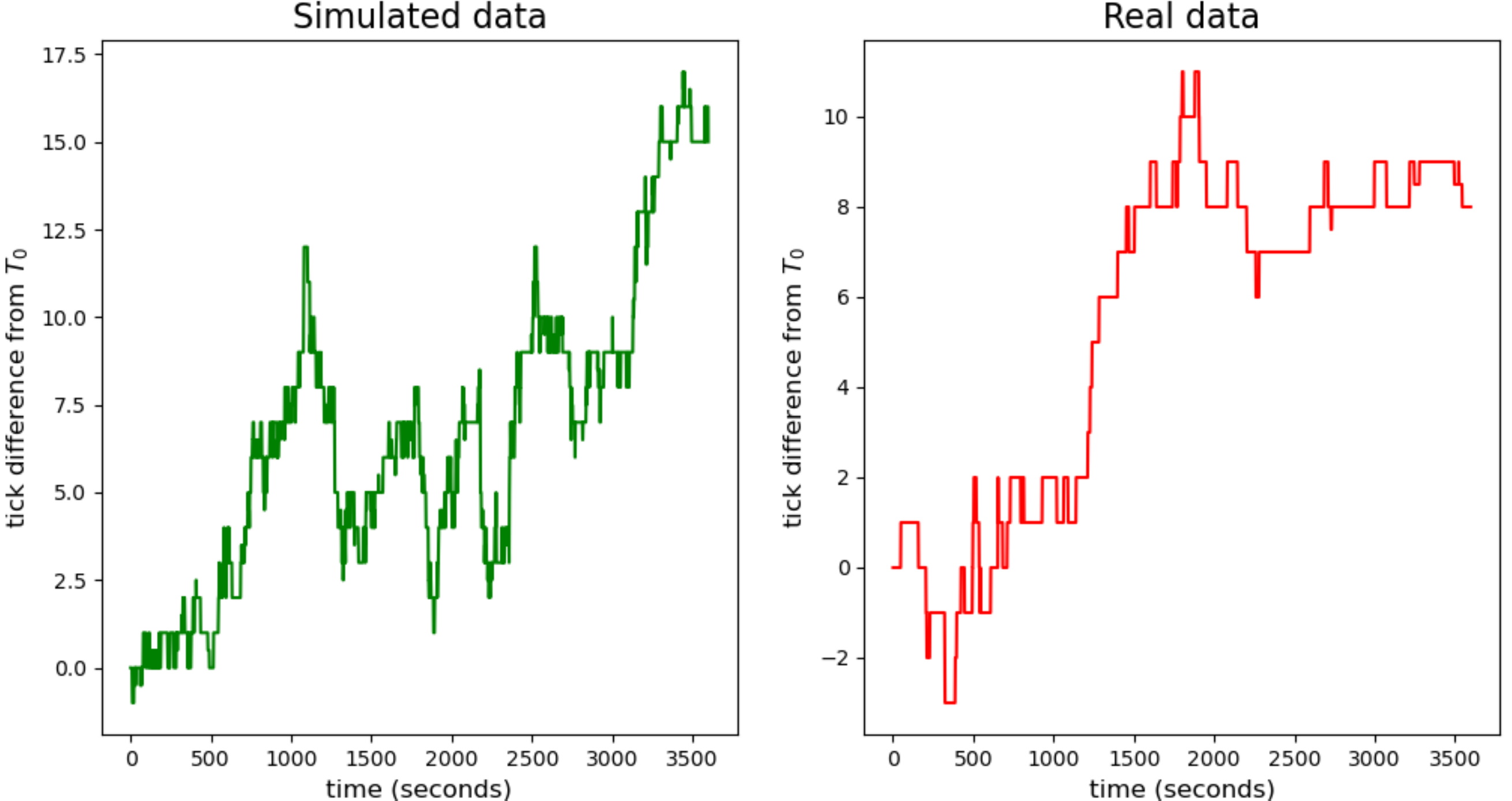}}
\hfil
\subfigure[Volatility-volume: Positive correlation of returns standard deviation and average traded volume in sim (left) and real (right) data.]{\label{fig:5}\includegraphics[width=0.4\linewidth]{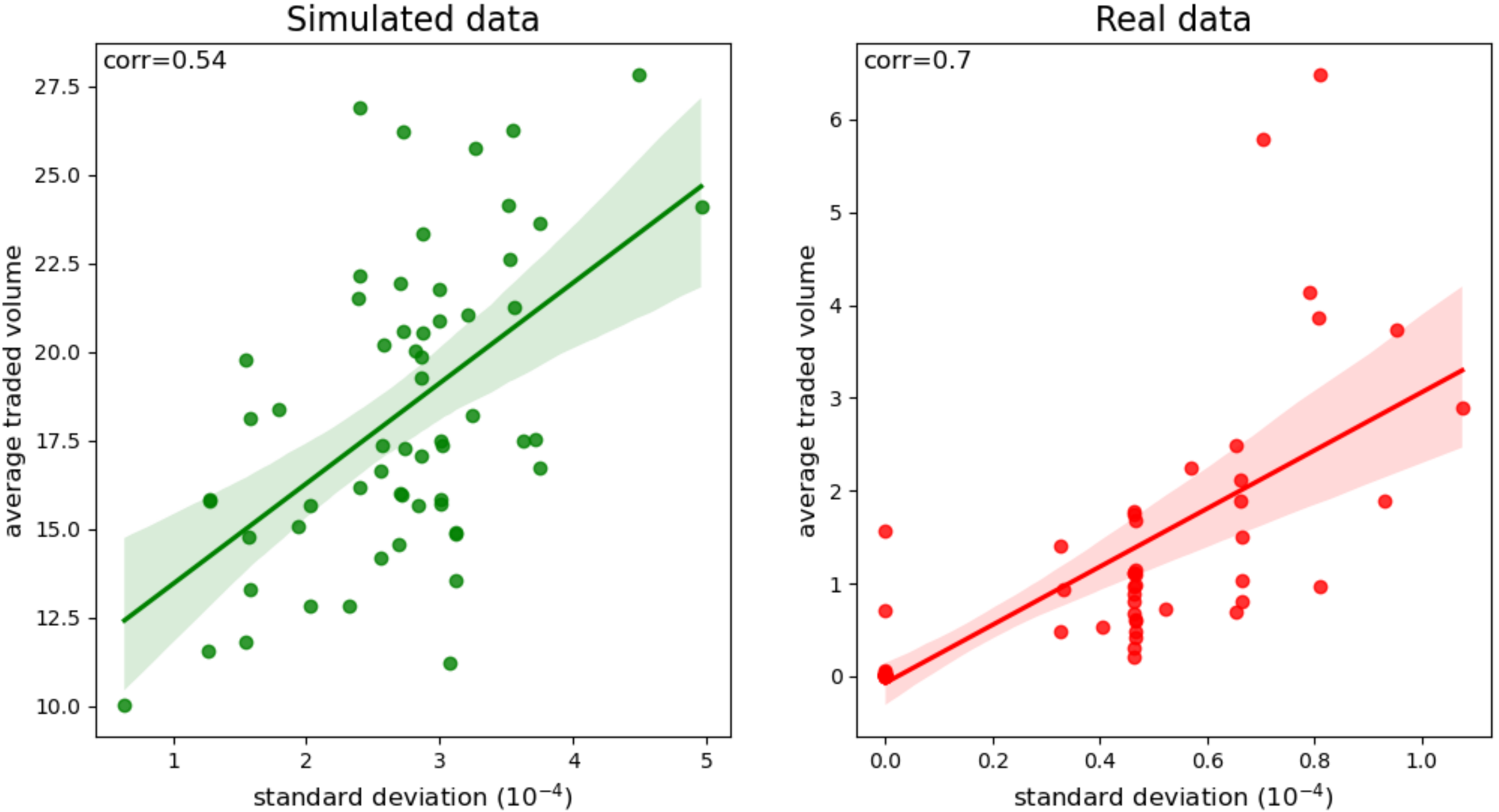}}
\medskip
\subfigure[Log returns: Log returns are normally distributed for sim (left) and real (right) data. As sampling frequency increases, kurtosis increases.]{\label{fig:3}\includegraphics[width=0.4\linewidth]{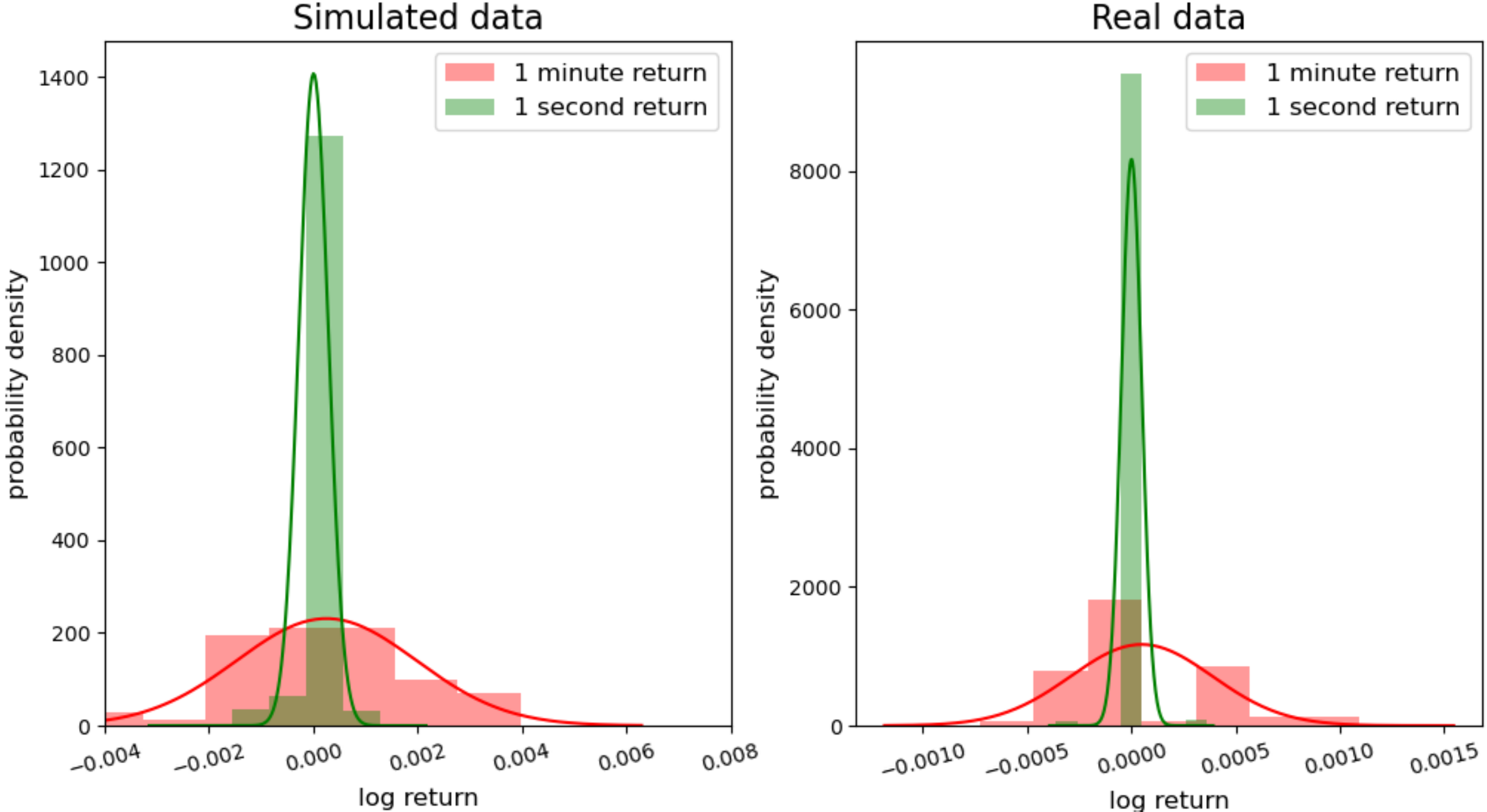}}
\hfil
\subfigure[Inter-arrival times: Exponentiated Weibull distribution is found to be the best fit for both simulated (left) and real (right) data.]{\label{fig:2}\includegraphics[width=0.4\linewidth]{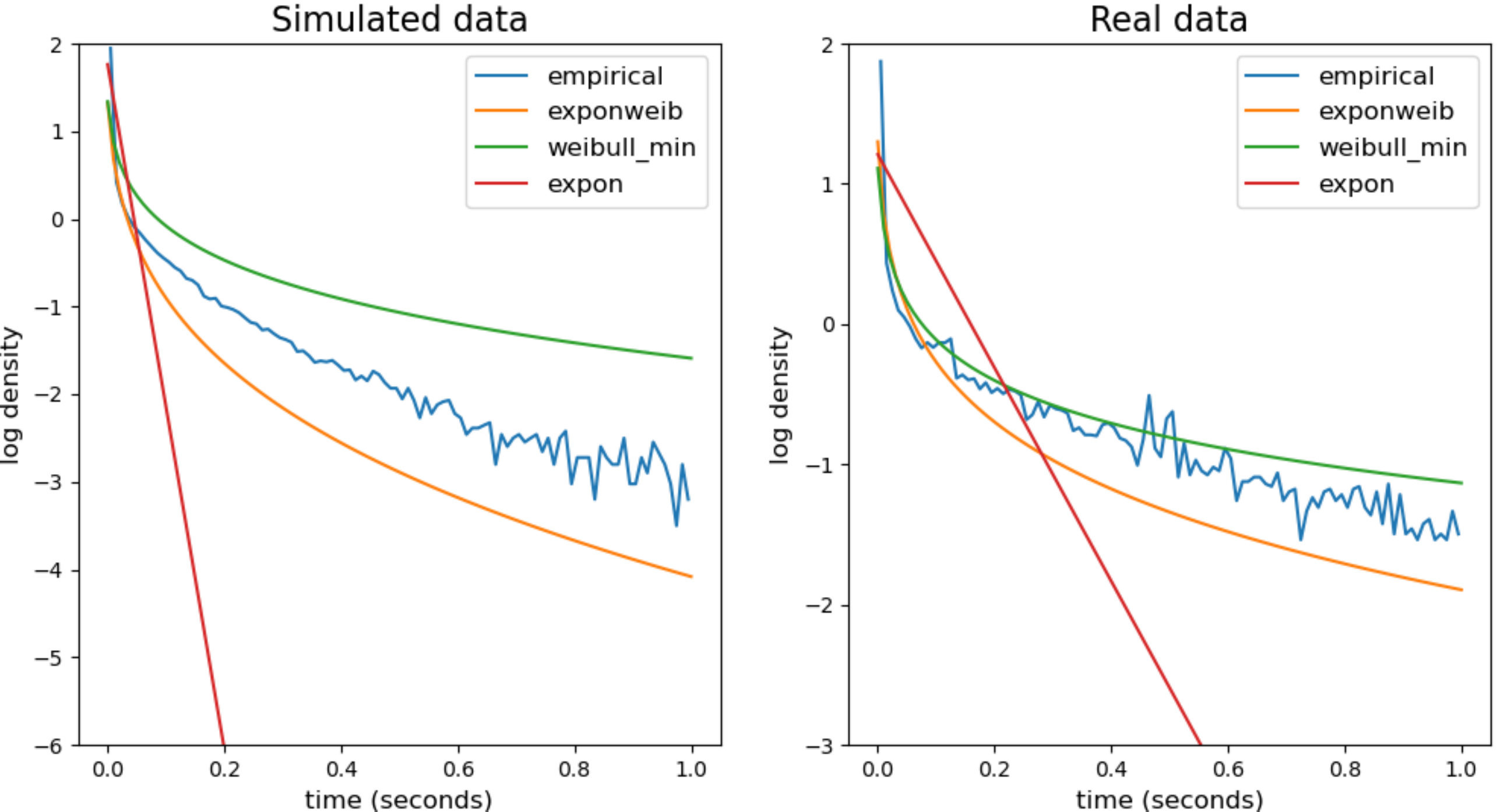}}
\medskip
\subfigure[Volatility clustering: sim (green); real (red). Correlation for absolute return over short time lags, decaying to zero as lag times increase.]{\label{fig:4}\includegraphics[width=0.4\linewidth]{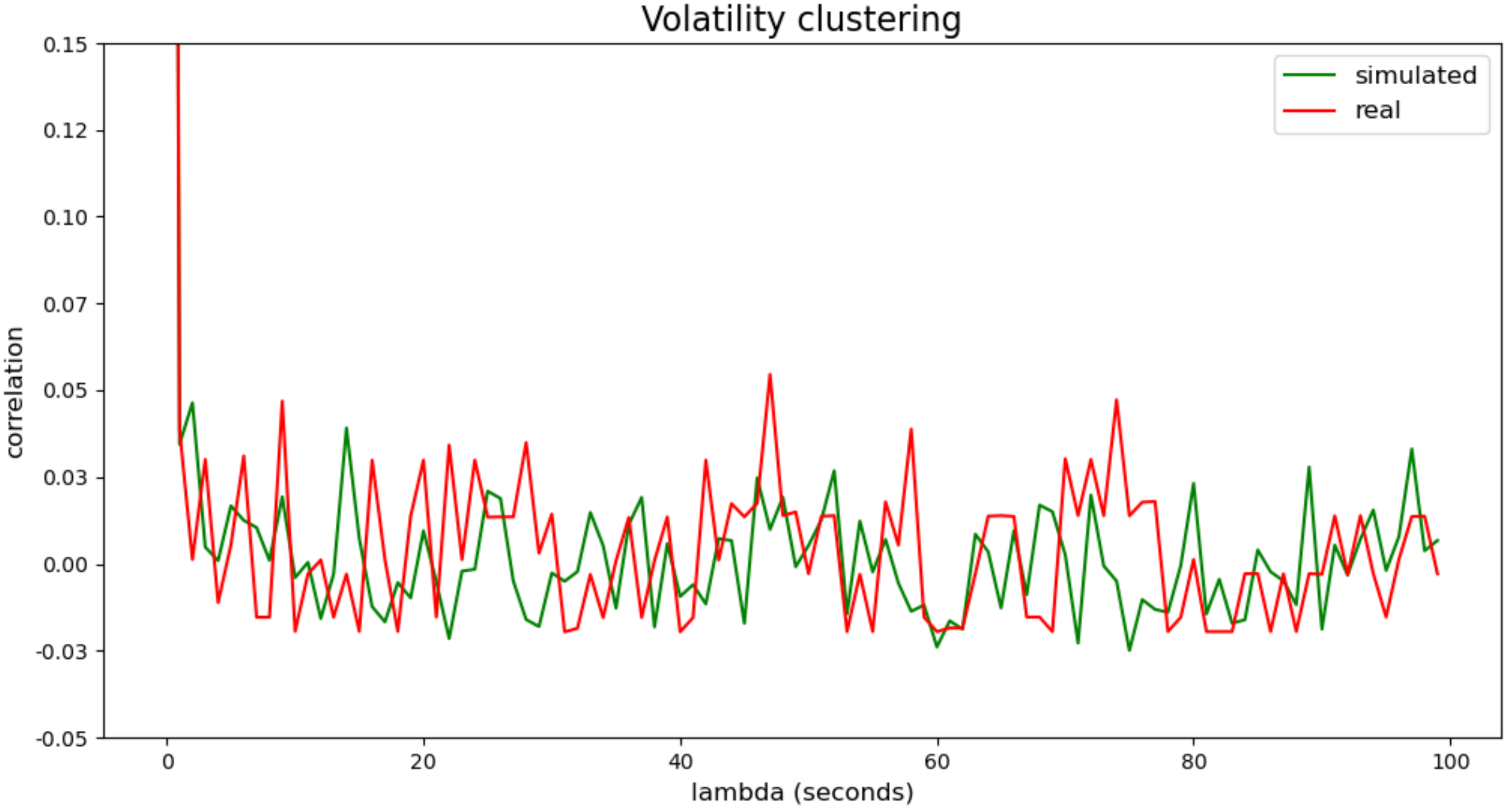}}
\hfil
\subfigure[Bid-ask spread: the spread is dominantly one tick over time.]{\label{fig:spread}\includegraphics[width=0.4\linewidth]{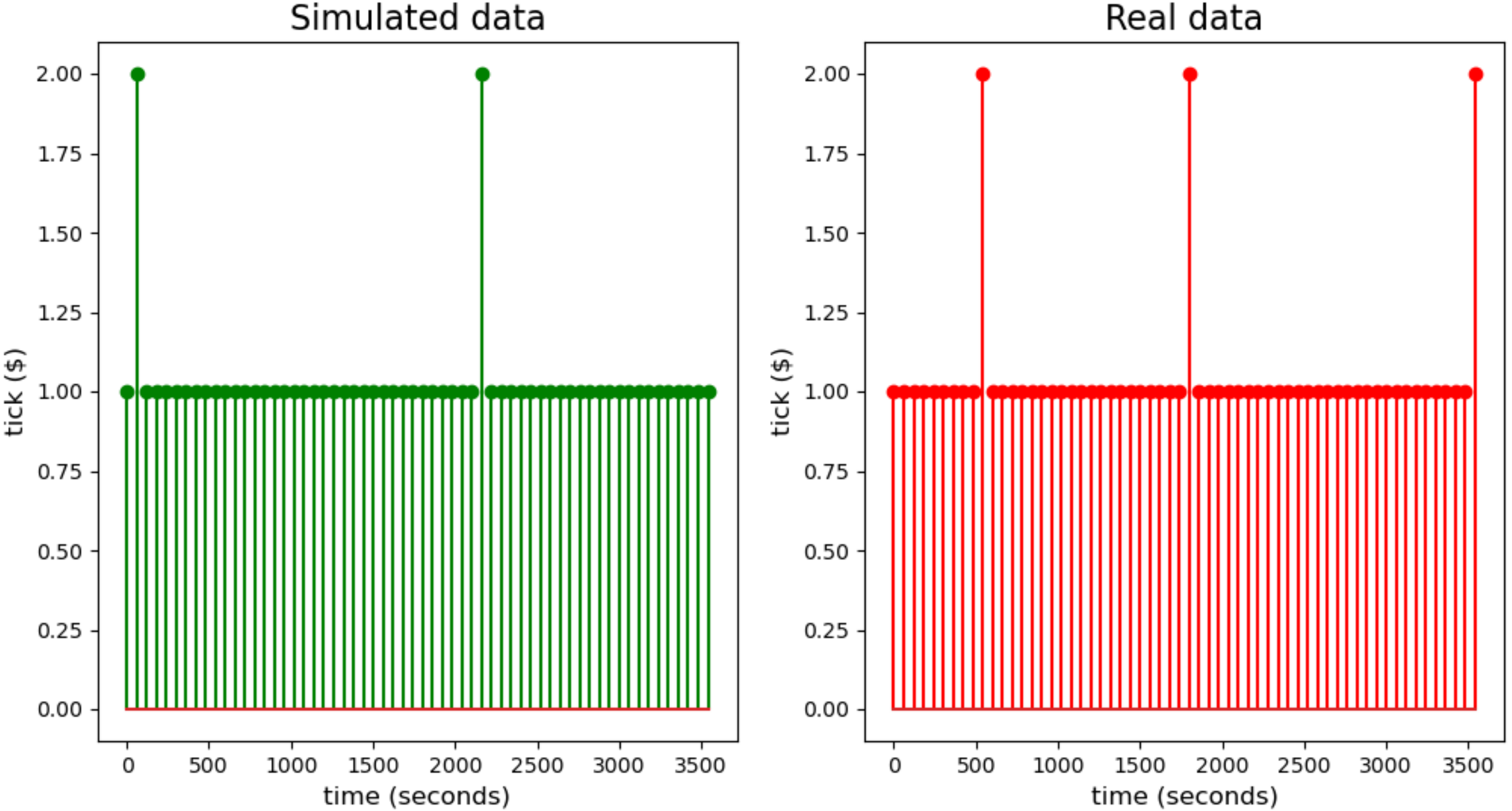}}
\medskip
\subfigure[Incoming order volume on quotes: both sim (left) data and
real (right) data show high and volatile incoming volume, with sim (left) showing higher mean and lower relative variation.]{\label{fig:involume}\includegraphics[width=0.4\linewidth]{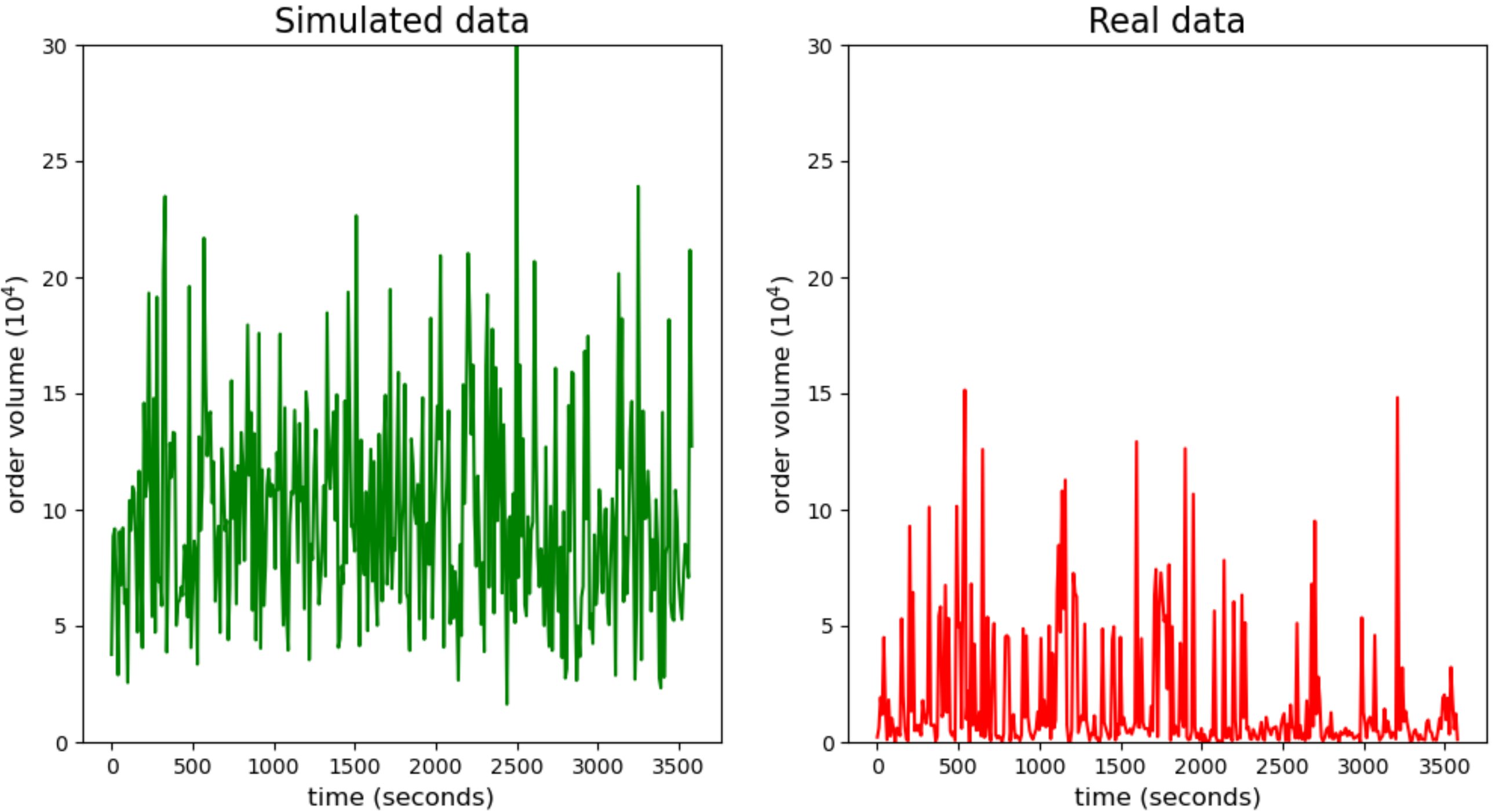}}
\hfil
\subfigure[Time to first fill power law distribution.]{\label{fig:ttff}\includegraphics[width=0.4\linewidth]{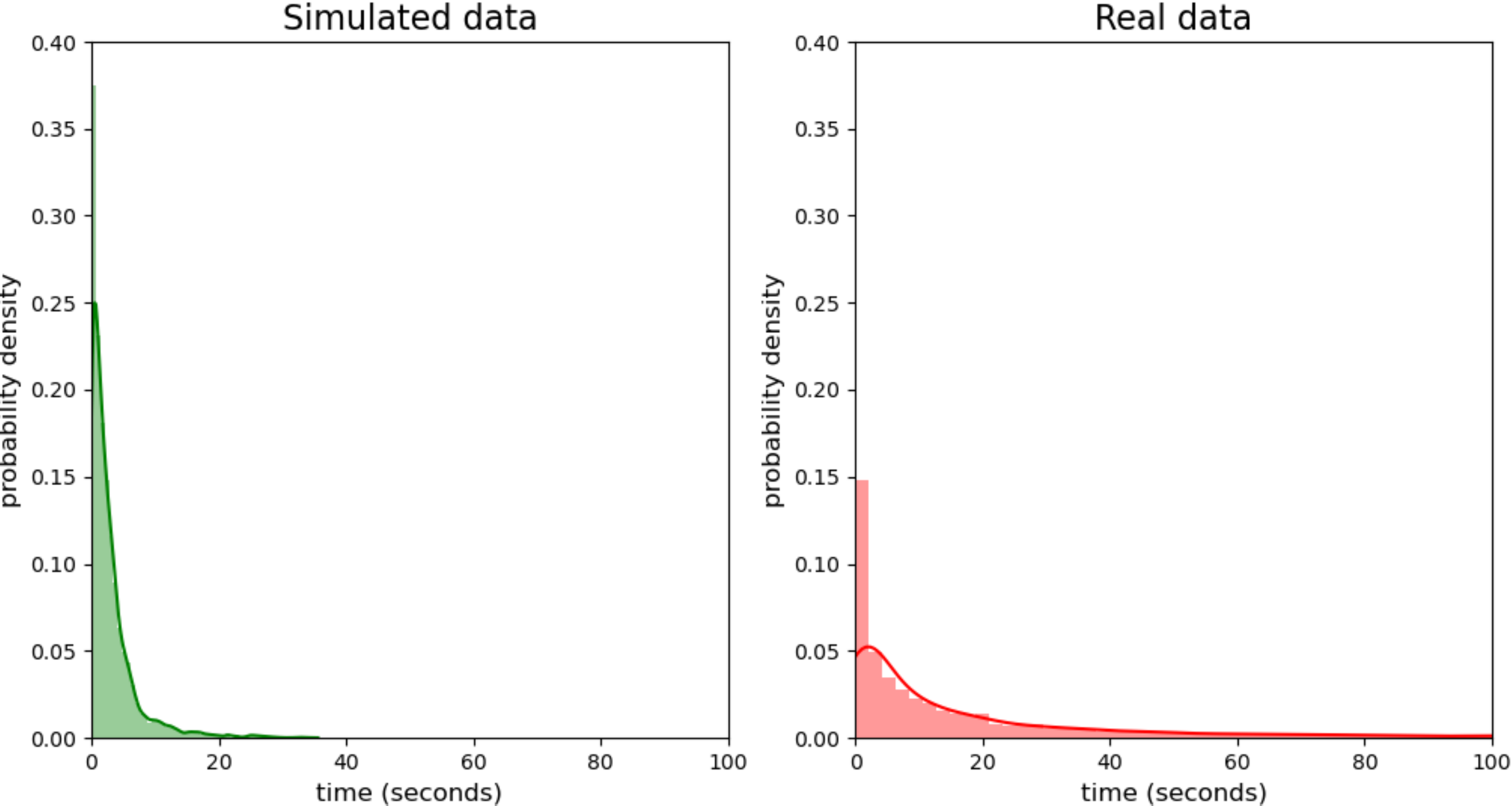}}
\caption{stylised facts and numerical properties verification for both simulated and real LOB data.}
\label{fig:images}
\end{figure*}

Facts previously verified in \cite{shi2022state}:
\begin{description}
    \item[Mid-price evolution.] This fact indicates that the evolution of the mid-price for liquid financial assets essentially follows a random walk and the mid-price remains volatile during any trading period \cite{li2020generating}. As shown in Fig.~\ref{fig:1}, the mid-price of both simulated and real data fluctuate over an approximately 10 minutes time interval. 
    
    \item[Autocorrelation in return series.] This fact, a reflection of the `efficient markets' hypothesis \cite{fama1991efficient}, indicates that price movements for liquid assets do not exhibit significant and strong autocorrelation. The absolute value for autocorrelation of log return time series $f(\tau)=\mathrm{corr}(r_{t+\tau,\Delta t},r_{t,\Delta t})$ was found to be lower than 0.1, and as time lag increases the coefficient converges to zero \cite{cont2001empirical}. Other studies found that the autocorrelation coefficient for return series to be weak but significant \cite{stanley2008statistical}. For simulated data, we find that the lag 1 autocorrelation to be around -0.1 and it fast decays as lag increases for one second frequency data. For real data, some samples show weak but significant autocorrelation, while others do not exhibit significant autocorrelation.

    \item[Normality of log returns.] This fact indicates that the distribution of asset log returns follows a normal distribution; at the same time, when sampling frequency changes from low to high, the kurtosis increases \cite{vyetrenko2020get}. Fig.~\ref{fig:3} shows the distribution of $r_{t,\Delta t}$ when the sampling frequency is high ($\Delta t=1 \;\mathrm{sec}$) and low ($\Delta t=1 \;\mathrm{min}$) can both be fitted with a Gaussian distribution. The distribution of high frequency return shows higher kurtosis and low tails ($\mathrm{kurtosis}>10$), while the low frequency return shows lower kurtosis and high tails ($\mathrm{kurtosis}< 1$). 
    
    \item[Event inter-arrival time.] This fact indicates that the distribution of event inter-arrival times follows an exponential or a Weibull distribution \cite{abergel2016limit}. Fig.~\ref{fig:2} presents the empirical density curve of time distribution (blue line) fitted with exponential (red), Weibull (green), and exponentiated Weibull (orange) distributions using MLE. Both simulated and real data have the best goodness of fit with the exponentiated Weibull distribution (the Jenson-Shannon divergence being 0.22 and 0.27 respectively).

    \item[Volatility/volume positive correlation.] This facts indicates that the standard deviation of log return $\sigma_{\tau,\Delta t}$ and traded volume $V_{\tau}$ have positive correlation, in the sense that trading activities of large volumes are likely to introduce higher volatility \cite{brandouy2012re}. In \cite{vyetrenko2020get}, the mean value for correlation coefficient for historical data lies around 0.4. Fig.~\ref{fig:5} plots the standard deviation against average trade volume on a scale of minutes. The correlation coefficient for simulated and real data are 0.57 and 0.72 respectively, which confirm the existence of this fact.
        
    \item[Volatility clustering.] This fact indicates that high volatility events tend to cluster in time. The function that used to quantify this feature is $f(\tau)=\mathrm{corr}(r_{t+\tau,\Delta t}^{2},r_{t,\Delta t}^{2})$ \cite{cont2001empirical}. This function remains positive and shows a downward tendency when delay in time $\tau$ increases. Fig.~\ref{fig:4} confirms this property is exhibited in both simulated and real data ($\Delta t=1 \;\mathrm{sec}$), and is similar to that shown in \cite{cont2001empirical}.
\end{description}

We were not able to reproduce the following stylised fact:

\begin{description}

    \item[Volatility/returns negative correlation.] This fact, also ter\-med as the `leverage effect' \cite{bouchaud2001more}, indicates that volatility and asset returns are negatively correlated. The volatility commonly used here is the implied volatility derived from a particular volatility index (e.g., VIX) or the price of financial derivatives. Here, there does not exist an implied volatility for the simulated asset. Therefore, following \cite{dufour2012measuring}, we investigate the relation between realised volatility (denoted as the sum of squared log returns) and returns. We find that both simulated and real data do not show a significant negative correlation between the two variables. This may suggest that realised volatility is not a good substitute of implied volatility.
    Although the leverage effect has previously been demonstrated using an ABM \cite{chen2013agent}, it is rare. In \cite{chen2013agent}, it was suggested that the reason many ABMs are not capable of recreating this fact is potentially due to the lack of asymmetric trading behaviours in simulation.

\end{description}

By taking reference from \cite{coletta2022learning}, we also exploit some numerical properties of the simulated LOB. Such properties vary a lot across different financial markets and financial assets, and they normally cannot be concluded with certain distributions with an agreed-upon parameter range. Hence, we compare the simulated data generated by the BT agent with the real data based on which the BT agent is trained, instead of conducting a direct comparison with \cite{coletta2022learning}.

\begin{description}
    
    \item [Bid-ask spread over time.] As both the simulated data and the real data is for highly liquid stocks that have comparatively low market value, the bid-ask spread is dominantly one tick over time, as shown in Fig.\ref{fig:spread}. Only when all orders resting on the top price levels are taken by market orders, or cancelled by agents, the bid-ask spread can be larger than one.
    
    \item [Incoming volume for limit order submissions.] Liquid fin\-ancial assets usually have high limit order incoming volume, and the volume fluctuates over time. From Fig.\ref{fig:involume} it can be seen both simulated and real data present high incoming volume and high volatility, with the simulated data exhibiting comparatively higher mean value and lower relative variation. The differences mainly originate from the setting of a stable distribution form for limit order submissions in the simulation, while in real markets the distributions of such properties can be dynamic over time according to market conditions and cannot be concluded with a stable form of distribution.
    
    \item [Time to first fill.] Time to first fill describes the time interval between the time when a order is submitted and the time when the order is partially fulfilled. This property indicates whether an asset is liquid or not, and liquid assets usually have low time to first fill. Here the majority of first time to fill for both simulated and real data are less than one second, as in Fig.\ref{fig:ttff}. Empirical research also indicated that the time to first fill follows a power law distribution \cite{abergel2016limit}, here both simulated and real data are in accordance with the literature (KS test p-value < 0.05)
    
\end{description}

\subsection{Experimental Agents Interaction}
\subsubsection{Interaction Experiment Settings}
By incorporating agents with various trading strategies into the ABIDES framework to interact with the BT, we can gain insights into how the system reacts to external stimuli, and how those reactions compare to real markets. Here, five criteria are considered: (i) mean profitability of agents (with no consideration of transaction cost); (ii) trading volume of agents as a proportion of total market trading volume; (iii) standard deviation of the realised mid-price time series; (iv) BT's ask-bid order imbalance: measured as $\mathrm{max}(AS+BC,BS+AC)/\mathrm{min}(AS+BC,BS+AC)-1$, in which $AS,AC,BS,BC$ indicate the number of ask submission, ask cancellation, bid submission and bid cancellation; and (v) the correlation between value agents' underlying fundamental stock value and realised stock price. 

We repeat each experimental condition 20 times and take mean values. Results are shown in Table~\ref{tab:stats}. We set number of agents $n\in\{1,15,50\}$, and set order flow impact {\em True} (i.e., BT reacts to agent orders). For control, we also compare markets with $n=(15)$ agents with order flow impact {\em False} (i.e., BT does not react to agent orders). 
Unless otherwise stated, agent populations are homogeneous.  
\subsubsection{ABIDES Configuration}

Four types of agents are considered: MM, MR, ZI, and HBL. To enable fair comparison, all agents wake up according to an exponential scheme (Poisson process), with average interval $\Delta t=30 secs$. The average order events posed by experimental agents as a percentage of all order events in the market is in the range $[0.1\%,5\%]$, ensuring that the market dynamics created by the BT is not overtaken by experimental agents. Other parameters are: $l_{1}=20$, $l_{2}=50$, $r_{max}=5$, $l=8$. For ABIDES related configurations, we refer to the Reference Market Simulation Configuration sample file. 
We set: mean value $\mu=10^{3}$, $\sigma^{2}=2 \times 10^{-10}$, reverting rate $\gamma=10^{-12}$. The original default values are: $10^{5}$, $10^{-8}$, $1.7\times10^{-16}$, respectively. We lower $\mu$ and $\sigma^{2}$ because we have lower starting asset value £10. We increase $\gamma$ so that the oracle demonstrates more mean reversion during 1-hour simulation window. Observation variance is set as $\sigma_{o}=10$. $R\_min=0$, $R\_max=5$, $eta=1$ are used to control the greediness of agents. Agents’ holding limit and starting cash are set big enough as that their decisions are not bounded by them. All findings to follow are supported by Wilcoxon signed-rank test (p-value<0.05, unless otherwise specified) either by comparing statistics in group $n=1$ and group $n=50$, or by comparing group $n=15$ and group $n=(15)$.

\begin{table*}[ht]
\caption{Agents and LOB statistics in markets containing one BT and $n$ homogeneous trading agents of type $T$. Where $n$ is shown in parentheses, experiments are conducted with no order flow impact. Criteria marked * are in $10^{-3}$.}
\label{tab:stats}
\small
\begin{tabular}{lcccccccccccccccccccc}
\toprule
\multicolumn{1}{c}{} & \multicolumn{4}{c}{(i) Profitability*} & \multicolumn{4}{c}{(ii) Proportion of trades*} & \multicolumn{4}{c}{(iii) Std of mid-price} & \multicolumn{4}{c}{(iv) Order imbalance*} & \multicolumn{4}{c}{(v) Correlation} \\ 
\cmidrule(rl){2-5} \cmidrule(rl){6-9} \cmidrule(rl){10-13} \cmidrule(rl){14-17} \cmidrule(rl){18-21} 
\diagbox{$T$}{$n$}
& 1 & 15 & (15) & 50 & 1 & 15 & (15) & 50 & 1 & 15 & (15) & 50 & 1 & 15 & (15) & 50 & 1 & 15 & (15) & 50\\ \midrule 
MM                                                                                        
& 23.1 & 43.1 & 22.7 & 91.1 & 2.0 & 25.2 & 27.8 & 67.2 & 13.4 & 24.6 & 14.0 & 47.1 & 6.9 & 19.6 & 4.6 & 55.5 & 0 & 0 & 0 & 0 \\
MR                                                                                     
& -15.2 & -7.9 & -20.3 & 0.4 & 2.3 & 34.0 & 29.9 & 113.8 & 12.2 & 8.1 & 10.9 & 3.9 & 7.9 & 15.1 & 7.9 & 27.2 & 0 & 0 & 0 & 0 \\
ZI                                                                                       
& -6.0 & -5.5 & -7.4 & -0.7 & 1.3 & 13.9 & 14.2 & 43.9 & 11.7 & 9.7 & 10.4 & 5.0 & 6.2 & 9.3 & 7.2 & 21.2 & 0.12 & 0.18 & -0.22 & 0.56 \\
HBL                      
& -4.2 & -4.1 & -6.3 & 0.8 & 1.0 & 13.1 & 13.1 & 44.2 & 13.4 & 9.4 & 9.6 & 4.1 & 6.3 & 11.0 & 8.4 & 26.2 & -0.04 & 0.06 & 0.06 & 0.58 \\ 
\bottomrule
\end{tabular}
\end{table*}

\subsubsection{Herding Effect}
Financial herding has been well documented in market empirical studies as an phenomenon of investors tending to follow the crowd or trend in the market instead of performing their own analysis \cite{zhou2009herding,boyd2016prevalence}. Here, making reference to Table~\ref{tab:stats}, we consider how a change in the number of agents holding the same trading strategy affects agent profitability and LOB behaviour. 

In terms of agent profitability (i), we see that mean profitability increases with number of agents in markets containing both trend and value strategies. Empirical studies on emerging and less-efficient markets indicate that the intensity of herding is positively related to trading profitability, especially for trend strategies \cite{chen2018contrarian,bikhchandani2000herd}. Studies that investigated the profitability of several technical trading strategies also revealed that by increasing the frequency of the data that the strategy is based on, the profitability of the trading strategy can be improved \cite{schulmeister2009profitability}. Also, as more agents adopt a particular trend strategy, such as momentum trading, there is greater influence on market price in a favourable (i.e., predictable) direction. Among all agents, MM achieves the highest profits. This profit is mainly derived from the capital gains in stock positions after causing an extreme one-direction price movement, as MM agents rarely neutralise a position after momentum ignition.

In terms of LOB volatility (iii), we see that increasing the number of MM agents causes increasing volatility, while the opposite effect occurs for other agent types. This is to be expected for momentum traders as price movements drive further movements in the same direction. Empirical studies of real markets indicate that herding behaviour is more intense in extreme market conditions, caused by momentum trading behaviours, and can lead to tail events such as a market crash \cite{boehmer2018competitive}. The authors of \cite{boehmer2018competitive} argued that homogeneous trading behaviours are more likely to cause price overreaction during a short time period, with deviation from the long term mean. In contrast, MR agents and value agents follow a mean-reverting fundamental value oracle. Therefore, an increase in the number of agents tends to reduce market volatility.

In terms of correlation between fundamental price and realised LOB price (v), empirical studies have indicated that the high intensity of HFTs' herding behaviour tends to increase the correlation of their order flows \cite{serrano2020high}, giving them more power to move the price. Studies on the futures market also indicated that HFTs help prices converge to the fundamental \cite{lee2015high}.
As value agents trade according to the comparison between their underlying fundamental value and the realised stock price, increasing the number of agents essentially gives the group more power to move price towards the fundamental value. We see the effect of this herding behaviour in the increased correlation for ZI and HBL. Notice that markets containing only trend agents exhibit zero correlation between the fundamental value and realised LOB price. This is to be expected as the fundamental value is ignored by these traders. 

\subsubsection{Order Flow Impact}
We can use the statistic of order imbalance generated by the BT agent (iv) to determine whether agents impact order flow. By comparing $n=15$ with the control $n=(15)$, we can clearly see that the degree of imbalance in the order stream is significantly larger when order flow impact is set to {\em True}. 

In markets dominated by trend agents, the order flow impact manifests as causing further trend following events (i.e., pushing price farther away, or pulling price back towards the mean). We can see this effect in the volatility of mid-price (iii). When no order flow impact is considered, volatility in a market full of MM traders tends to be underestimated; and the same statistic in a market full of MR traders tends to be overestimated. The underestimation of volatility in a MM market is caused by overlooking the empirical finding of momentum ignition \cite{biais2014hft}. Momentum ignition indicates that investors tend to follow the price trend made by HFTs. Following a similar logic, the overestimated volatility in a MR market also results from the exclusion of market reactions to mean reverting orders.

Studies of real markets indicate that large institutional `value' traders contribute substantially to price discovery \cite{nawn2019price}. We can observe a similar effect in the simulation results. We see that an increase in the number of value agents in the market causes an increase in the correlation between realised LOB price and the fundamental value (v). Also, correlation is smaller in the control, where order flow impact is excluded. This result indicates that the inclusion of order flow impact can help model the price discovery role of value agents.

\subsubsection{Competition between agent strategies}
To understand strategy interaction, we also performed heterogeneous experiments, with the existence of order flow impact: (1) `trend' markets containing 15 MM $vs$ 15 MR; and (2) `value' markets containing 15 ZI $vs$ 15 HBL. In trend markets, we find that the profits of both strategies are inferior to the profits generated in homogeneous markets, as these two types of agents are competing with each other to influence the price in their respective favor. Also, the resulting market volatility falls by nearly 50\% compared with a homogeneous MM market, indicating MR agents' mean-reverting impact. In value markets, profits of both strategies are not significantly different with their respective homogeneous markets. However, in relative terms, HBL remains  more profitable than ZI. This is unsurprising given HBL has a relatively sophisticated trading strategy, when compared with ZI. 

\subsubsection{Stylised facts with interaction}\label{sec:stylisedfactswithinteraction}
Finally, we measure the sty\-lised facts of simulated markets where BT interacts with agents via order flow impact. We consider homogeneous markets containing each trader type, with $n=15$ and order flow impact set {\em True}. Results are: (1) Hurst exponent: $[0.48,0.65]$; (2) Autocorrelation of order signs: $[0.24,0.26]$ for submission and $[0.17,0.18]$ for cancellation; (3) OFI R-squared: $[0.54,0.71]$; and (4) Price impact function slope: $[0.18,0.48]$. In addition, the list of six stylised facts previously verified in \cite{shi2022state} are also exhibited by the model.   

\begin{figure}[t]
\centering
\subfigure[Mid price difference when $\lambda=0.01$]{\label{fig:pov1}\includegraphics[width=1\linewidth]{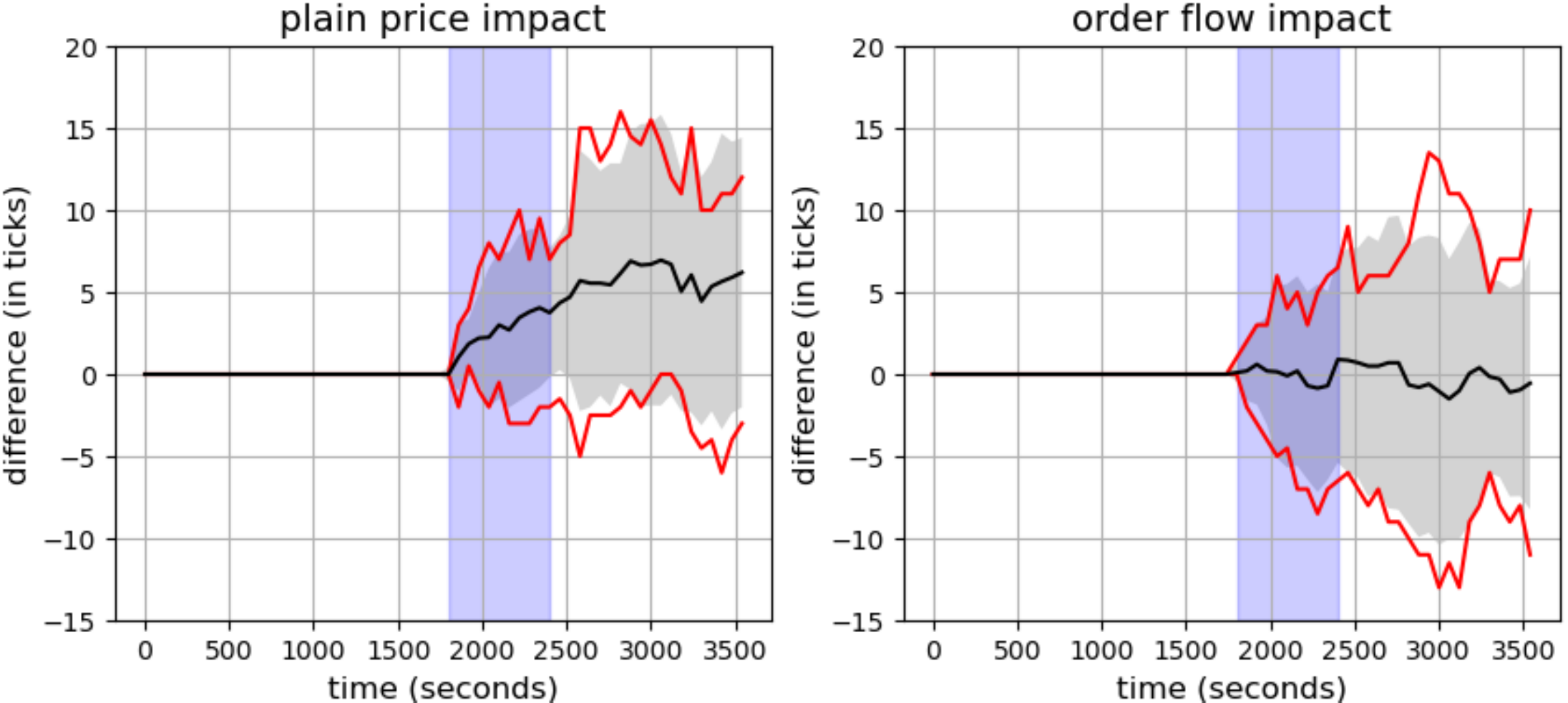}}
\medskip
\subfigure[Mid price difference when $\lambda=0.1$]{\label{fig:pov2}\includegraphics[width=1\linewidth]{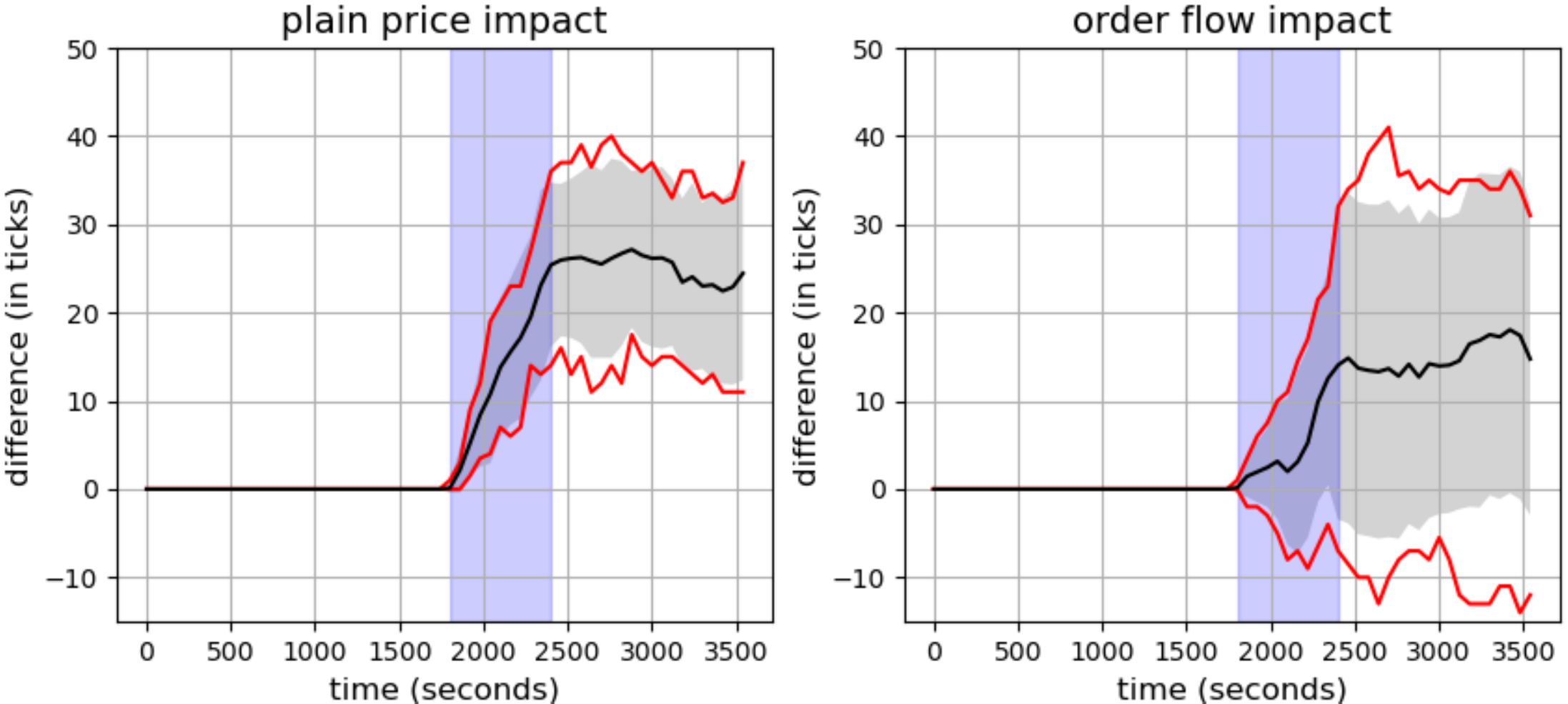}}
\medskip
\subfigure[Mid price difference when $\lambda=0.2$]{\label{fig:pov3}\includegraphics[width=1\linewidth]{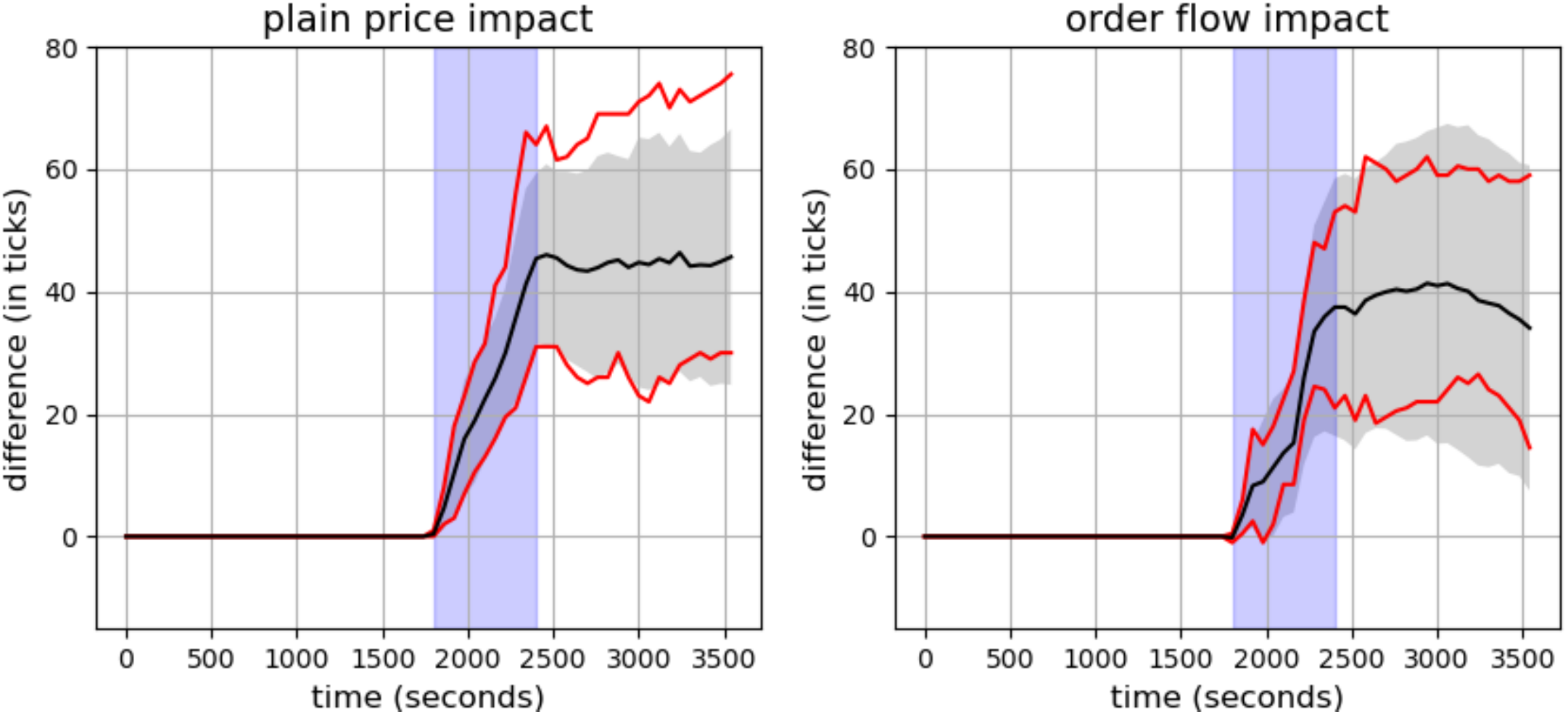}}
\medskip
\subfigure[Mid price difference when $\lambda=0.5$]{\label{fig:pov4}\includegraphics[width=1\linewidth]{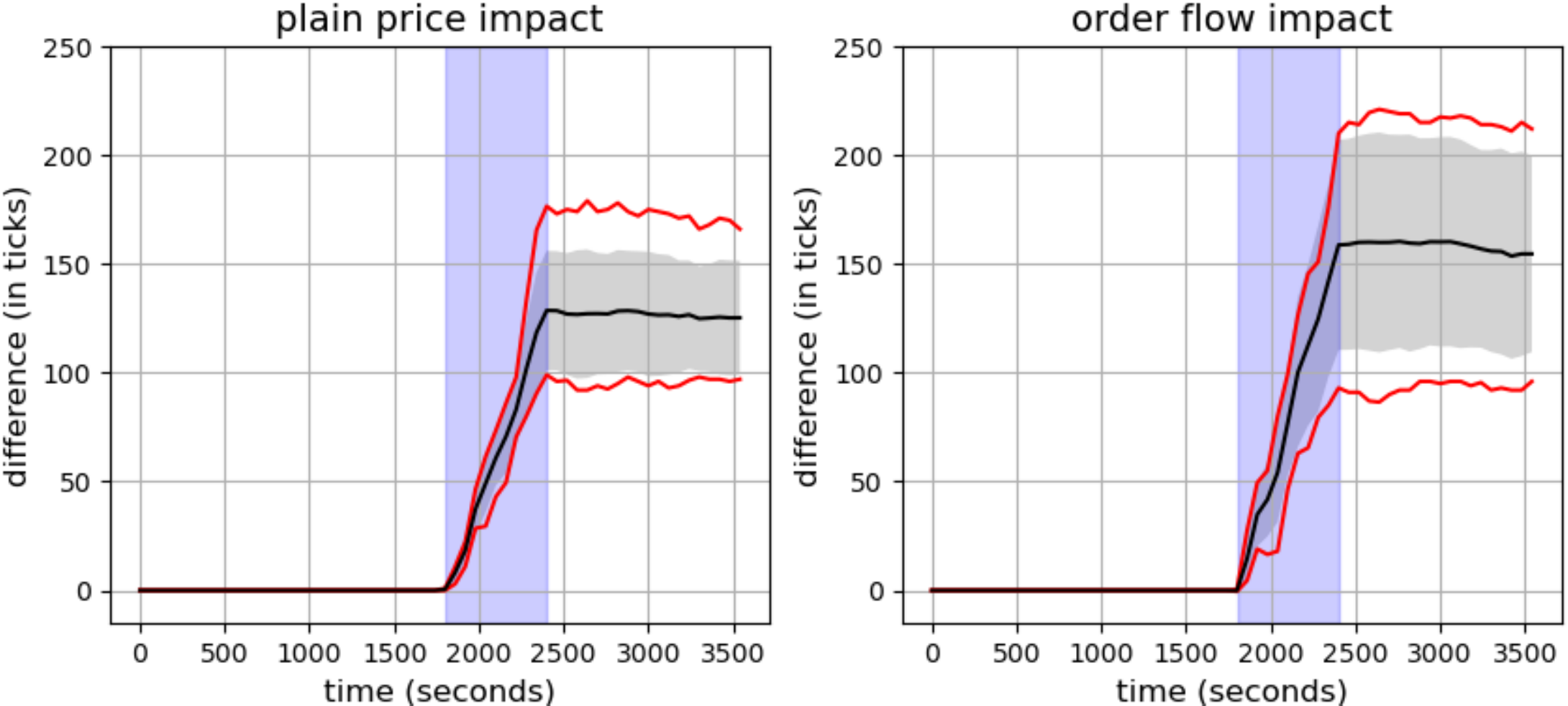}}
\hfil
\caption{The influence of plain price impact and order flow impact on the asset price as percentage of volume $\lambda$ varies.}
\label{fig:pov}
\end{figure}

\subsection{Responsiveness of the system}
Here, we closely follow the approach of \cite{coletta2022learning} to test the responsiveness of the system to price impact. We introduce a percent-of-volume (POV) agent which simulates a large volume trader that is likely to move the market price.
The POV agent is defined by two parameters: time interval $T$ seconds, and percentage of market volume $\lambda$. At time $t$, POV is instructed to buy/sell a total volume $V=\lambda M$, where M is the total transaction volume of the whole market over the previous $T$ seconds. POV will attempt to trade volume $V$ over the next $T$ seconds by splitting $V$ into multiple smaller orders. 

We run experiments for one simulated hour and set POV to begin trading at time $t=1800$ seconds, with trading interval $T=600$ seconds; i.e., POV will do nothing for the initial 30 minutes, followed by a burst of trading between 30-40 minutes, and then no further trading until the simulation ends. We explore the price impact of increasing trade volumes, such that $\lambda \in [0.01,0.1,0.2,0.5]$.

For each set of experiments, we run the simulation under three conditions: (1) A market containing BT agent only. Here there is no POV agent and so we are able to trace the evolution of the market when there is no impact; (2) A market containing POV agent and BT agent configured to have no order flow impact such that the BT does not react to the exogenous orders posted by the POV. Here, any price change in the market is a direct result of POV order submissions only; (3) A market containing POV agent and BT agent configured to have order flow impact. In this system, changes in market price result from both the POV order submissions and also the responsive market behaviours of the BT. 

By comparing the difference in market price between configurations (1) and (2), we are able to investigate the price impact of simply adding orders into the market and we refer to this price difference as `plain' price impact. This impact resembles traditional backtesting, where orders eat the book and are not replaced as the market cannot respond. 
By comparing the difference in market price between configurations (2) and (3), we can investigate how the market responds to POV order volume. 
We refer to this `extra' price impact as `order flow' impact. 
This approach of separating total price impact into two components is unique, and different to the approach taken by \cite{coletta2022learning}.

Results are shown in Fig.\ref{fig:pov}, with the left hand side showing `plain' impact calculated as the difference in mid-price generated in (1) and (2), and the right hand side showing `order flow' impact calculated as the difference in mid-price generated in (2) and (3). The black line, red line, and grey shaded area are mean value, 10-th and 90-th percentiles, and one standard deviation from the mean, respectively. The POV agent submits orders during the blue shaded area. Each simulation is repeated 10 times, and the POV agent only submits bid orders.

For plain price impact, we see that price rises when POV trades, and grows monotonically with $\lambda$. Price impact is permanent and remains after the POV agent stops trading. 
For order flow impact, we see that low values of $\lambda=0.01$ do not cause the market to respond. As $\lambda$ increases, order flow impact increases superlinearly. This shows the market adversely responding to increased buy pressure, pushing market prices higher than they would otherwise go.

\section{Conclusions and Future Works}
We have presented the hybrid NS-ABM model for realistic LOB simulation and implemented the model using the ABIDES framework \cite{byrd2020abides}. NS-ABM combines the benefits of ABM with data-driven approaches to simulation by including a neural stochastic BT that is pre-trained on real data. The BT trader has been shown to realistically simulate real world LOB dynamics, with ten stylised facts -- empirically observed properties of markets that are accepted as fact -- approximately reproduced (see Section~\ref{sec:stylisedfacts}). In addition, the BT trader can also realistically react to endogenous market events, with stylised facts remaining once a populations of trend and value trading agents are added to the simulation (see Section~\ref{sec:stylisedfactswithinteraction}). Since the NS-ABM model can realistically replicate market characteristics, it removes the need to include populations of stochastic `noise' agents that approximately characterise market dynamics (e.g., \cite{mcgroarty2019high}). NS-ABM can also act as a `dynamic back-test' harness, such that individual trading strategies can be evaluated on historical data that realistically adapts to the actions of the trading strategy. 

In the experiments we performed, an exogeneous fundamental value was generated using a mean-reverting stochastic process. However, ABIDES enables historical data to be used as a fundamental value. Using such a configuration, the NS-ABM offers a potential route towards the `holy grail' of dynamic back-testing for financial trading algorithms, where trading events generated by a strategy under test have trading impact. It would also be interesting to conduct more diversified interaction experiments, for instance simulations that include more heterogeneous trading strategies, to see how strategies interfere with each other and how the resultant market statistics revolve. In terms of the BT, we are also eager to investigate the feasibility of replicating structural changes (e.g. a market crash) that embedded in the training data, which provides the possibility of replicating more realistic, or even abnormal, market dynamics in LOB simulation. We intend to explore this exciting avenue of investigation in future work.



\begin{acks}
    ZS's PhD is funded by a China Scholarship Council / University of Bristol joint-scholarship. 
\end{acks}


\ifnum\ISLONG=1{\clearpage}\else{}\fi
\balance
\bibliographystyle{ACM-Reference-Format} 
\bibliography{aamas}


\begin{thebibliography}{51}


\ifx \showCODEN    \undefined \def \showCODEN     #1{\unskip}     \fi
\ifx \showDOI      \undefined \def \showDOI       #1{#1}\fi
\ifx \showISBNx    \undefined \def \showISBNx     #1{\unskip}     \fi
\ifx \showISBNxiii \undefined \def \showISBNxiii  #1{\unskip}     \fi
\ifx \showISSN     \undefined \def \showISSN      #1{\unskip}     \fi
\ifx \showLCCN     \undefined \def \showLCCN      #1{\unskip}     \fi
\ifx \shownote     \undefined \def \shownote      #1{#1}          \fi
\ifx \showarticletitle \undefined \def \showarticletitle #1{#1}   \fi
\ifx \showURL      \undefined \def \showURL       {\relax}        \fi
\providecommand\bibfield[2]{#2}
\providecommand\bibinfo[2]{#2}
\providecommand\natexlab[1]{#1}
\providecommand\showeprint[2][]{arXiv:#2}

\bibitem[\protect\citeauthoryear{Abergel, Anane, Chakraborti, Jedidi, and
  Toke}{Abergel et~al\mbox{.}}{2016}]%
        {abergel2016limit}
\bibfield{author}{\bibinfo{person}{Fr{\'e}d{\'e}ric Abergel},
  \bibinfo{person}{Marouane Anane}, \bibinfo{person}{Anirban Chakraborti},
  \bibinfo{person}{Aymen Jedidi}, {and} \bibinfo{person}{Ioane~Muni Toke}.}
  \bibinfo{year}{2016}\natexlab{}.
\newblock \bibinfo{booktitle}{\emph{Limit order books}}.
\newblock \bibinfo{publisher}{Cambridge University Press},
  \bibinfo{address}{Cambridge, UK}.
\newblock


\bibitem[\protect\citeauthoryear{Biais, Foucault, et~al\mbox{.}}{Biais
  et~al\mbox{.}}{2014}]%
        {biais2014hft}
\bibfield{author}{\bibinfo{person}{Bruno Biais}, \bibinfo{person}{Thierry
  Foucault}, {et~al\mbox{.}}} \bibinfo{year}{2014}\natexlab{}.
\newblock \showarticletitle{HFT and market quality}.
\newblock \bibinfo{journal}{\emph{Bankers, Markets \& Investors}}
  \bibinfo{volume}{128}, \bibinfo{number}{1} (\bibinfo{year}{2014}),
  \bibinfo{pages}{5--19}.
\newblock


\bibitem[\protect\citeauthoryear{Bikhchandani and Sharma}{Bikhchandani and
  Sharma}{2000}]%
        {bikhchandani2000herd}
\bibfield{author}{\bibinfo{person}{Sushil Bikhchandani} {and}
  \bibinfo{person}{Sunil Sharma}.} \bibinfo{year}{2000}\natexlab{}.
\newblock \showarticletitle{Herd behavior in financial markets}.
\newblock \bibinfo{journal}{\emph{IMF Staff papers}} \bibinfo{volume}{47},
  \bibinfo{number}{3} (\bibinfo{year}{2000}), \bibinfo{pages}{279--310}.
\newblock


\bibitem[\protect\citeauthoryear{Boehmer, Li, and Saar}{Boehmer
  et~al\mbox{.}}{2018}]%
        {boehmer2018competitive}
\bibfield{author}{\bibinfo{person}{Ekkehart Boehmer}, \bibinfo{person}{Dan Li},
  {and} \bibinfo{person}{Gideon Saar}.} \bibinfo{year}{2018}\natexlab{}.
\newblock \showarticletitle{The competitive landscape of high-frequency trading
  firms}.
\newblock \bibinfo{journal}{\emph{The Review of Financial Studies}}
  \bibinfo{volume}{31}, \bibinfo{number}{6} (\bibinfo{year}{2018}),
  \bibinfo{pages}{2227--2276}.
\newblock


\bibitem[\protect\citeauthoryear{Bouchaud, M{\'e}zard, and Potters}{Bouchaud
  et~al\mbox{.}}{2002}]%
        {bouchaud2002statistical}
\bibfield{author}{\bibinfo{person}{Jean-Philippe Bouchaud},
  \bibinfo{person}{Marc M{\'e}zard}, {and} \bibinfo{person}{Marc Potters}.}
  \bibinfo{year}{2002}\natexlab{}.
\newblock \showarticletitle{Statistical properties of stock order books:
  empirical results and models}.
\newblock \bibinfo{journal}{\emph{Quantitative finance}} \bibinfo{volume}{2},
  \bibinfo{number}{4} (\bibinfo{year}{2002}), \bibinfo{pages}{251}.
\newblock


\bibitem[\protect\citeauthoryear{Bouchaud and Potters}{Bouchaud and
  Potters}{2001}]%
        {bouchaud2001more}
\bibfield{author}{\bibinfo{person}{Jean-Philippe Bouchaud} {and}
  \bibinfo{person}{Marc Potters}.} \bibinfo{year}{2001}\natexlab{}.
\newblock \showarticletitle{More stylized facts of financial markets: leverage
  effect and downside correlations}.
\newblock \bibinfo{journal}{\emph{Physica A: Statistical Mechanics and its
  Applications}} \bibinfo{volume}{299}, \bibinfo{number}{1-2}
  (\bibinfo{year}{2001}), \bibinfo{pages}{60--70}.
\newblock


\bibitem[\protect\citeauthoryear{Bouchaud and Potters}{Bouchaud and
  Potters}{2003}]%
        {bouchaud2003theory}
\bibfield{author}{\bibinfo{person}{Jean-Philippe Bouchaud} {and}
  \bibinfo{person}{Marc Potters}.} \bibinfo{year}{2003}\natexlab{}.
\newblock \bibinfo{booktitle}{\emph{Theory of Financial Risk and Derivative
  Pricing: From Statistical Physics to Risk Management (2nd ed.)}}.
\newblock \bibinfo{publisher}{Cambridge university press},
  \bibinfo{address}{Cambridge}.
\newblock


\bibitem[\protect\citeauthoryear{Boyd, B{\"u}y{\"u}k{\c{s}}ahin, Haigh, and
  Harris}{Boyd et~al\mbox{.}}{2016}]%
        {boyd2016prevalence}
\bibfield{author}{\bibinfo{person}{Naomi~E Boyd}, \bibinfo{person}{Bahattin
  B{\"u}y{\"u}k{\c{s}}ahin}, \bibinfo{person}{Michael~S Haigh}, {and}
  \bibinfo{person}{Jeffrey~H Harris}.} \bibinfo{year}{2016}\natexlab{}.
\newblock \showarticletitle{The prevalence, sources, and effects of herding}.
\newblock \bibinfo{journal}{\emph{Journal of Futures Markets}}
  \bibinfo{volume}{36}, \bibinfo{number}{7} (\bibinfo{year}{2016}),
  \bibinfo{pages}{671--694}.
\newblock


\bibitem[\protect\citeauthoryear{Brandouy, Corelli, Veryzhenko, and
  Waldeck}{Brandouy et~al\mbox{.}}{2012}]%
        {brandouy2012re}
\bibfield{author}{\bibinfo{person}{Olivier Brandouy}, \bibinfo{person}{Angelo
  Corelli}, \bibinfo{person}{Iryna Veryzhenko}, {and} \bibinfo{person}{Roger
  Waldeck}.} \bibinfo{year}{2012}\natexlab{}.
\newblock \showarticletitle{A re-examination of the “zero is enough”
  hypothesis in the emergence of financial stylized facts}.
\newblock \bibinfo{journal}{\emph{Journal of Economic Interaction and
  Coordination}} \bibinfo{volume}{7}, \bibinfo{number}{2}
  (\bibinfo{year}{2012}), \bibinfo{pages}{223--248}.
\newblock


\bibitem[\protect\citeauthoryear{Byrd}{Byrd}{2019}]%
        {byrd2019explaining}
\bibfield{author}{\bibinfo{person}{David Byrd}.}
  \bibinfo{year}{2019}\natexlab{}.
\newblock \bibinfo{title}{Explaining agent-based financial market simulation}.
\newblock \bibinfo{howpublished}{{arXiv:1909.11650v1}}.
\newblock
\urldef\tempurl%
\url{https://doi.org/10.48550/arXiv.1909.11650}
\showDOI{\tempurl}


\bibitem[\protect\citeauthoryear{Byrd, Hybinette, and Balch}{Byrd
  et~al\mbox{.}}{2020}]%
        {byrd2020abides}
\bibfield{author}{\bibinfo{person}{David Byrd}, \bibinfo{person}{Maria
  Hybinette}, {and} \bibinfo{person}{Tucker~Hybinette Balch}.}
  \bibinfo{year}{2020}\natexlab{}.
\newblock \showarticletitle{ABIDES: Towards High-Fidelity Multi-Agent Market
  Simulation}. In \bibinfo{booktitle}{\emph{Proceedings of the 2020 ACM SIGSIM
  Conference on Principles of Advanced Discrete Simulation}} (Miami, FL, Spain)
  \emph{(\bibinfo{series}{SIGSIM-PADS '20})}. \bibinfo{publisher}{Association
  for Computing Machinery}, \bibinfo{address}{New York, NY, USA},
  \bibinfo{pages}{11–22}.
\newblock
\showISBNx{9781450375924}


\bibitem[\protect\citeauthoryear{Chen, Zheng, and Tan}{Chen
  et~al\mbox{.}}{2013}]%
        {chen2013agent}
\bibfield{author}{\bibinfo{person}{Jun-Jie Chen}, \bibinfo{person}{Bo Zheng},
  {and} \bibinfo{person}{Lei Tan}.} \bibinfo{year}{2013}\natexlab{}.
\newblock \showarticletitle{Agent-based model with asymmetric trading and
  herding for complex financial systems}.
\newblock \bibinfo{journal}{\emph{PloS one}} \bibinfo{volume}{8},
  \bibinfo{number}{11} (\bibinfo{year}{2013}), \bibinfo{pages}{e79531}.
\newblock


\bibitem[\protect\citeauthoryear{Chen, Hua, and Jiang}{Chen
  et~al\mbox{.}}{2018}]%
        {chen2018contrarian}
\bibfield{author}{\bibinfo{person}{Qiwei Chen}, \bibinfo{person}{Xiuping Hua},
  {and} \bibinfo{person}{Ying Jiang}.} \bibinfo{year}{2018}\natexlab{}.
\newblock \showarticletitle{Contrarian strategy and herding behaviour in the
  Chinese stock market}.
\newblock \bibinfo{journal}{\emph{The European Journal of Finance}}
  \bibinfo{volume}{24}, \bibinfo{number}{16} (\bibinfo{year}{2018}),
  \bibinfo{pages}{1552--1568}.
\newblock


\bibitem[\protect\citeauthoryear{Coletta, Moulin, Vyetrenko, and Balch}{Coletta
  et~al\mbox{.}}{2022}]%
        {coletta2022learning}
\bibfield{author}{\bibinfo{person}{Andrea Coletta}, \bibinfo{person}{Aymeric
  Moulin}, \bibinfo{person}{Svitlana Vyetrenko}, {and} \bibinfo{person}{Tucker
  Balch}.} \bibinfo{year}{2022}\natexlab{}.
\newblock \showarticletitle{Learning to simulate realistic limit order book
  markets from data as a World Agent}. In \bibinfo{booktitle}{\emph{Proceedings
  of the Third ACM International Conference on AI in Finance}}.
  \bibinfo{publisher}{Association for Computing Machinery},
  \bibinfo{address}{New York, NY, USA}, \bibinfo{pages}{428--436}.
\newblock


\bibitem[\protect\citeauthoryear{Cont}{Cont}{2001}]%
        {cont2001empirical}
\bibfield{author}{\bibinfo{person}{Rama Cont}.}
  \bibinfo{year}{2001}\natexlab{}.
\newblock \showarticletitle{Empirical properties of asset returns: stylized
  facts and statistical issues}.
\newblock \bibinfo{journal}{\emph{Quantitative finance}} \bibinfo{volume}{1},
  \bibinfo{number}{2} (\bibinfo{year}{2001}), \bibinfo{pages}{223}.
\newblock


\bibitem[\protect\citeauthoryear{Cont}{Cont}{2007}]%
        {cont2007volatility}
\bibfield{author}{\bibinfo{person}{Rama Cont}.}
  \bibinfo{year}{2007}\natexlab{}.
\newblock \showarticletitle{Volatility clustering in financial markets:
  empirical facts and agent-based models}.
\newblock In \bibinfo{booktitle}{\emph{Long memory in economics}},
  \bibfield{editor}{\bibinfo{person}{Gilles Teyssière} {and}
  \bibinfo{person}{Alan~P. Kirman}} (Eds.). \bibinfo{publisher}{Springer},
  \bibinfo{address}{Berlin, Heidelberg}, \bibinfo{pages}{289--309}.
\newblock


\bibitem[\protect\citeauthoryear{Cont and De~Larrard}{Cont and
  De~Larrard}{2013}]%
        {cont2013price}
\bibfield{author}{\bibinfo{person}{Rama Cont} {and} \bibinfo{person}{Adrien
  De~Larrard}.} \bibinfo{year}{2013}\natexlab{}.
\newblock \showarticletitle{Price dynamics in a Markovian limit order market}.
\newblock \bibinfo{journal}{\emph{SIAM Journal on Financial Mathematics}}
  \bibinfo{volume}{4}, \bibinfo{number}{1} (\bibinfo{year}{2013}),
  \bibinfo{pages}{1--25}.
\newblock


\bibitem[\protect\citeauthoryear{Cont, Kukanov, and Stoikov}{Cont
  et~al\mbox{.}}{2014}]%
        {cont2014price}
\bibfield{author}{\bibinfo{person}{Rama Cont}, \bibinfo{person}{Arseniy
  Kukanov}, {and} \bibinfo{person}{Sasha Stoikov}.}
  \bibinfo{year}{2014}\natexlab{}.
\newblock \showarticletitle{The price impact of order book events}.
\newblock \bibinfo{journal}{\emph{Journal of financial econometrics}}
  \bibinfo{volume}{12}, \bibinfo{number}{1} (\bibinfo{year}{2014}),
  \bibinfo{pages}{47--88}.
\newblock


\bibitem[\protect\citeauthoryear{Cont, Stoikov, and Talreja}{Cont
  et~al\mbox{.}}{2010}]%
        {cont2010stochastic}
\bibfield{author}{\bibinfo{person}{Rama Cont}, \bibinfo{person}{Sasha Stoikov},
  {and} \bibinfo{person}{Rishi Talreja}.} \bibinfo{year}{2010}\natexlab{}.
\newblock \showarticletitle{A stochastic model for order book dynamics}.
\newblock \bibinfo{journal}{\emph{Operations research}} \bibinfo{volume}{58},
  \bibinfo{number}{3} (\bibinfo{year}{2010}), \bibinfo{pages}{549--563}.
\newblock


\bibitem[\protect\citeauthoryear{Duffin and Cartlidge}{Duffin and
  Cartlidge}{2018}]%
        {duffin18}
\bibfield{author}{\bibinfo{person}{Matthew Duffin} {and} \bibinfo{person}{John
  Cartlidge}.} \bibinfo{year}{2018}\natexlab{}.
\newblock \showarticletitle{Agent-Based Model Exploration of Latency Arbitrage
  in Fragmented Financial Markets}. In \bibinfo{booktitle}{\emph{2018 IEEE
  Symposium Series on Computational Intelligence (SSCI)}}.
  \bibinfo{publisher}{IEEE}, \bibinfo{address}{New York, NY, USA},
  \bibinfo{pages}{2312--2320}.
\newblock


\bibitem[\protect\citeauthoryear{Dufour, Garcia, and Taamouti}{Dufour
  et~al\mbox{.}}{2012}]%
        {dufour2012measuring}
\bibfield{author}{\bibinfo{person}{Jean-Marie Dufour},
  \bibinfo{person}{Ren{\'e} Garcia}, {and} \bibinfo{person}{Abderrahim
  Taamouti}.} \bibinfo{year}{2012}\natexlab{}.
\newblock \showarticletitle{Measuring high-frequency causality between returns,
  realized volatility, and implied volatility}.
\newblock \bibinfo{journal}{\emph{Journal of Financial Econometrics}}
  \bibinfo{volume}{10}, \bibinfo{number}{1} (\bibinfo{year}{2012}),
  \bibinfo{pages}{124--163}.
\newblock


\bibitem[\protect\citeauthoryear{Fama}{Fama}{1991}]%
        {fama1991efficient}
\bibfield{author}{\bibinfo{person}{Eugene~F Fama}.}
  \bibinfo{year}{1991}\natexlab{}.
\newblock \showarticletitle{Efficient capital markets: II}.
\newblock \bibinfo{journal}{\emph{The journal of finance}}
  \bibinfo{volume}{46}, \bibinfo{number}{5} (\bibinfo{year}{1991}),
  \bibinfo{pages}{1575--1617}.
\newblock


\bibitem[\protect\citeauthoryear{Feng, Li, Podobnik, Preis, and Stanley}{Feng
  et~al\mbox{.}}{2012}]%
        {feng2012linking}
\bibfield{author}{\bibinfo{person}{Ling Feng}, \bibinfo{person}{Baowen Li},
  \bibinfo{person}{Boris Podobnik}, \bibinfo{person}{Tobias Preis}, {and}
  \bibinfo{person}{H~Eugene Stanley}.} \bibinfo{year}{2012}\natexlab{}.
\newblock \showarticletitle{Linking agent-based models and stochastic models of
  financial markets}.
\newblock \bibinfo{journal}{\emph{Proceedings of the National Academy of
  Sciences}} \bibinfo{volume}{109}, \bibinfo{number}{22}
  (\bibinfo{year}{2012}), \bibinfo{pages}{8388--8393}.
\newblock


\bibitem[\protect\citeauthoryear{Gould, Porter, Williams, McDonald, Fenn, and
  Howison}{Gould et~al\mbox{.}}{2013}]%
        {gould2013limit}
\bibfield{author}{\bibinfo{person}{Martin~D Gould}, \bibinfo{person}{Mason~A
  Porter}, \bibinfo{person}{Stacy Williams}, \bibinfo{person}{Mark McDonald},
  \bibinfo{person}{Daniel~J Fenn}, {and} \bibinfo{person}{Sam~D Howison}.}
  \bibinfo{year}{2013}\natexlab{}.
\newblock \showarticletitle{Limit order books}.
\newblock \bibinfo{journal}{\emph{Quantitative Finance}} \bibinfo{volume}{13},
  \bibinfo{number}{11} (\bibinfo{year}{2013}), \bibinfo{pages}{1709--1742}.
\newblock


\bibitem[\protect\citeauthoryear{Gu and Zhou}{Gu and Zhou}{2009}]%
        {gu2009emergence}
\bibfield{author}{\bibinfo{person}{Gao-Feng Gu} {and} \bibinfo{person}{Wei-Xing
  Zhou}.} \bibinfo{year}{2009}\natexlab{}.
\newblock \showarticletitle{Emergence of long memory in stock volatility from a
  modified Mike-Farmer model}.
\newblock \bibinfo{journal}{\emph{EPL (Europhysics Letters)}}
  \bibinfo{volume}{86}, \bibinfo{number}{4} (\bibinfo{year}{2009}),
  \bibinfo{pages}{48002}.
\newblock


\bibitem[\protect\citeauthoryear{Karpe, Fang, Ma, and Wang}{Karpe
  et~al\mbox{.}}{2020}]%
        {karpe2020multi}
\bibfield{author}{\bibinfo{person}{Micha{\"e}l Karpe}, \bibinfo{person}{Jin
  Fang}, \bibinfo{person}{Zhongyao Ma}, {and} \bibinfo{person}{Chen Wang}.}
  \bibinfo{year}{2020}\natexlab{}.
\newblock \showarticletitle{Multi-agent reinforcement learning in a realistic
  limit order book market simulation}. In \bibinfo{booktitle}{\emph{Proceedings
  of the First ACM International Conference on AI in Finance}}.
  \bibinfo{publisher}{Association for Computing Machinery},
  \bibinfo{address}{New York, NY, USA}, \bibinfo{pages}{1--7}.
\newblock


\bibitem[\protect\citeauthoryear{Kumar}{Kumar}{2021}]%
        {kumar2021deep}
\bibfield{author}{\bibinfo{person}{Pankaj Kumar}.}
  \bibinfo{year}{2021}\natexlab{}.
\newblock \bibinfo{title}{Deep Hawkes Process for High-Frequency Market
  Making}.
\newblock
\newblock
\urldef\tempurl%
\url{https://doi.org/10.48550/ARXIV.2109.15110}
\showDOI{\tempurl}


\bibitem[\protect\citeauthoryear{Lee}{Lee}{2015}]%
        {lee2015high}
\bibfield{author}{\bibinfo{person}{Eun~Jung Lee}.}
  \bibinfo{year}{2015}\natexlab{}.
\newblock \showarticletitle{High frequency trading in the Korean index futures
  market}.
\newblock \bibinfo{journal}{\emph{Journal of Futures Markets}}
  \bibinfo{volume}{35}, \bibinfo{number}{1} (\bibinfo{year}{2015}),
  \bibinfo{pages}{31--51}.
\newblock


\bibitem[\protect\citeauthoryear{Li, Wang, Lin, Sinha, and Wellman}{Li
  et~al\mbox{.}}{2020}]%
        {li2020generating}
\bibfield{author}{\bibinfo{person}{Junyi Li}, \bibinfo{person}{Xintong Wang},
  \bibinfo{person}{Yaoyang Lin}, \bibinfo{person}{Arunesh Sinha}, {and}
  \bibinfo{person}{Michael Wellman}.} \bibinfo{year}{2020}\natexlab{}.
\newblock \showarticletitle{Generating realistic stock market order streams}.
  In \bibinfo{booktitle}{\emph{Proceedings of the AAAI Conference on Artificial
  Intelligence}}, Vol.~\bibinfo{volume}{34(01)}. \bibinfo{publisher}{AAAI
  Press}, \bibinfo{address}{Palo Alto, CA}, \bibinfo{pages}{727--734}.
\newblock


\bibitem[\protect\citeauthoryear{Lillo and Farmer}{Lillo and Farmer}{2004}]%
        {lillo2004long}
\bibfield{author}{\bibinfo{person}{Fabrizio Lillo} {and}
  \bibinfo{person}{J~Doyne Farmer}.} \bibinfo{year}{2004}\natexlab{}.
\newblock \showarticletitle{The long memory of the efficient market}.
\newblock \bibinfo{journal}{\emph{Studies in nonlinear dynamics \&
  econometrics}} \bibinfo{volume}{8}, \bibinfo{number}{3}
  (\bibinfo{year}{2004}), \bibinfo{pages}{1--33}.
\newblock


\bibitem[\protect\citeauthoryear{Lillo, Farmer, and Mantegna}{Lillo
  et~al\mbox{.}}{2003}]%
        {lillo2003master}
\bibfield{author}{\bibinfo{person}{Fabrizio Lillo}, \bibinfo{person}{J~Doyne
  Farmer}, {and} \bibinfo{person}{Rosario~N Mantegna}.}
  \bibinfo{year}{2003}\natexlab{}.
\newblock \showarticletitle{Master curve for price-impact function}.
\newblock \bibinfo{journal}{\emph{Nature}} \bibinfo{volume}{421},
  \bibinfo{number}{6919} (\bibinfo{year}{2003}), \bibinfo{pages}{129--130}.
\newblock


\bibitem[\protect\citeauthoryear{McGroarty, Booth, Gerding, and
  Chinthalapati}{McGroarty et~al\mbox{.}}{2019}]%
        {mcgroarty2019high}
\bibfield{author}{\bibinfo{person}{Frank McGroarty}, \bibinfo{person}{Ash
  Booth}, \bibinfo{person}{Enrico Gerding}, {and} \bibinfo{person}{VL
  Chinthalapati}.} \bibinfo{year}{2019}\natexlab{}.
\newblock \showarticletitle{High frequency trading strategies, market fragility
  and price spikes: an agent based model perspective}.
\newblock \bibinfo{journal}{\emph{Annals of Operations Research}}
  \bibinfo{volume}{282}, \bibinfo{number}{1} (\bibinfo{year}{2019}),
  \bibinfo{pages}{217--244}.
\newblock


\bibitem[\protect\citeauthoryear{N{\ae}s and Skjeltorp}{N{\ae}s and
  Skjeltorp}{2006}]%
        {naes2006order}
\bibfield{author}{\bibinfo{person}{Randi N{\ae}s} {and}
  \bibinfo{person}{Johannes~A Skjeltorp}.} \bibinfo{year}{2006}\natexlab{}.
\newblock \showarticletitle{Order book characteristics and the
  volume--volatility relation: Empirical evidence from a limit order market}.
\newblock \bibinfo{journal}{\emph{Journal of Financial Markets}}
  \bibinfo{volume}{9}, \bibinfo{number}{4} (\bibinfo{year}{2006}),
  \bibinfo{pages}{408--432}.
\newblock


\bibitem[\protect\citeauthoryear{Nawn and Banerjee}{Nawn and Banerjee}{2019}]%
        {nawn2019price}
\bibfield{author}{\bibinfo{person}{Samarpan Nawn} {and} \bibinfo{person}{Ashok
  Banerjee}.} \bibinfo{year}{2019}\natexlab{}.
\newblock \showarticletitle{Do the limit orders of proprietary and agency
  algorithmic traders discover or obscure security prices?}
\newblock \bibinfo{journal}{\emph{Journal of Empirical Finance}}
  \bibinfo{volume}{53} (\bibinfo{year}{2019}), \bibinfo{pages}{109--125}.
\newblock
\urldef\tempurl%
\url{https://doi.org/10.1016/j.jempfin.2019.06.003}
\showDOI{\tempurl}


\bibitem[\protect\citeauthoryear{Nolte, Salmon, and Adcock}{Nolte
  et~al\mbox{.}}{2016}]%
        {nolte2016high}
\bibfield{author}{\bibinfo{person}{Ingmar Nolte}, \bibinfo{person}{Mark
  Salmon}, {and} \bibinfo{person}{Chris Adcock}.}
  \bibinfo{year}{2016}\natexlab{}.
\newblock \bibinfo{booktitle}{\emph{High frequency trading and limit order book
  dynamics}}.
\newblock \bibinfo{publisher}{Routledge}, \bibinfo{address}{Abingdon, England}.
\newblock


\bibitem[\protect\citeauthoryear{Paddrik, Hayes, Todd, Yang, Beling, and
  Scherer}{Paddrik et~al\mbox{.}}{2012}]%
        {paddrik2012agent}
\bibfield{author}{\bibinfo{person}{Mark Paddrik}, \bibinfo{person}{Roy Hayes},
  \bibinfo{person}{Andrew Todd}, \bibinfo{person}{Steve Yang},
  \bibinfo{person}{Peter Beling}, {and} \bibinfo{person}{William Scherer}.}
  \bibinfo{year}{2012}\natexlab{}.
\newblock \showarticletitle{An agent based model of the E-Mini S\&P 500 applied
  to Flash Crash analysis}. In \bibinfo{booktitle}{\emph{2012 IEEE Conference
  on Computational Intelligence for Financial Engineering \& Economics
  (CIFEr)}}. \bibinfo{publisher}{IEEE}, \bibinfo{address}{New York, NY, USA},
  \bibinfo{pages}{1--8}.
\newblock


\bibitem[\protect\citeauthoryear{Panayi and Peters}{Panayi and Peters}{2015}]%
        {panayi2015stochastic}
\bibfield{author}{\bibinfo{person}{Efstathios Panayi} {and}
  \bibinfo{person}{Gareth~W Peters}.} \bibinfo{year}{2015}\natexlab{}.
\newblock \showarticletitle{Stochastic simulation framework for the limit order
  book using liquidity-motivated agents}.
\newblock \bibinfo{journal}{\emph{International Journal of Financial
  Engineering}} \bibinfo{volume}{2}, \bibinfo{number}{02}
  (\bibinfo{year}{2015}), \bibinfo{pages}{1550013}.
\newblock


\bibitem[\protect\citeauthoryear{Peng, Buldyrev, Havlin, Simons, Stanley, and
  Goldberger}{Peng et~al\mbox{.}}{1994}]%
        {peng1994mosaic}
\bibfield{author}{\bibinfo{person}{C-K Peng}, \bibinfo{person}{Sergey~V
  Buldyrev}, \bibinfo{person}{Shlomo Havlin}, \bibinfo{person}{Michael Simons},
  \bibinfo{person}{H~Eugene Stanley}, {and} \bibinfo{person}{Ary~L
  Goldberger}.} \bibinfo{year}{1994}\natexlab{}.
\newblock \showarticletitle{Mosaic organization of DNA nucleotides}.
\newblock \bibinfo{journal}{\emph{Physical review e}} \bibinfo{volume}{49},
  \bibinfo{number}{2} (\bibinfo{year}{1994}), \bibinfo{pages}{1685}.
\newblock


\bibitem[\protect\citeauthoryear{Preis, Golke, Paul, and Schneider}{Preis
  et~al\mbox{.}}{2007}]%
        {preis2007statistical}
\bibfield{author}{\bibinfo{person}{Tobias Preis}, \bibinfo{person}{Sebastian
  Golke}, \bibinfo{person}{Wolfgang Paul}, {and} \bibinfo{person}{Johannes~J
  Schneider}.} \bibinfo{year}{2007}\natexlab{}.
\newblock \showarticletitle{Statistical analysis of financial returns for a
  multiagent order book model of asset trading}.
\newblock \bibinfo{journal}{\emph{Physical Review E}} \bibinfo{volume}{76},
  \bibinfo{number}{1} (\bibinfo{year}{2007}), \bibinfo{pages}{016108}.
\newblock


\bibitem[\protect\citeauthoryear{Schulmeister}{Schulmeister}{2009}]%
        {schulmeister2009profitability}
\bibfield{author}{\bibinfo{person}{Stephan Schulmeister}.}
  \bibinfo{year}{2009}\natexlab{}.
\newblock \showarticletitle{Profitability of technical stock trading: Has it
  moved from daily to intraday data?}
\newblock \bibinfo{journal}{\emph{Review of Financial Economics}}
  \bibinfo{volume}{18}, \bibinfo{number}{4} (\bibinfo{year}{2009}),
  \bibinfo{pages}{190--201}.
\newblock


\bibitem[\protect\citeauthoryear{Serrano}{Serrano}{2020}]%
        {serrano2020high}
\bibfield{author}{\bibinfo{person}{Antonio~S{\'a}nchez Serrano}.}
  \bibinfo{year}{2020}\natexlab{}.
\newblock \showarticletitle{High-frequency trading and systemic risk: A
  structured review of findings and policies}.
\newblock \bibinfo{journal}{\emph{Review of Economics}} \bibinfo{volume}{71},
  \bibinfo{number}{3} (\bibinfo{year}{2020}), \bibinfo{pages}{169--195}.
\newblock


\bibitem[\protect\citeauthoryear{Shi and Cartlidge}{Shi and Cartlidge}{2021}]%
        {shi2021limit}
\bibfield{author}{\bibinfo{person}{Zijian Shi} {and} \bibinfo{person}{John
  Cartlidge}.} \bibinfo{year}{2021}\natexlab{}.
\newblock \showarticletitle{The Limit Order Book Recreation Model (LOBRM): An
  Extended Analysis}. In \bibinfo{booktitle}{\emph{Joint European Conference on
  Machine Learning and Knowledge Discovery in Databases}}.
  \bibinfo{publisher}{Springer}, \bibinfo{address}{Cham, Switzerland},
  \bibinfo{pages}{204--220}.
\newblock


\bibitem[\protect\citeauthoryear{Shi and Cartlidge}{Shi and Cartlidge}{2022}]%
        {shi2022state}
\bibfield{author}{\bibinfo{person}{Zijian Shi} {and} \bibinfo{person}{John
  Cartlidge}.} \bibinfo{year}{2022}\natexlab{}.
\newblock \showarticletitle{State Dependent Parallel Neural Hawkes Process for
  Limit Order Book Event Stream Prediction and Simulation}. In
  \bibinfo{booktitle}{\emph{Proceedings of the 28th ACM SIGKDD Conference on
  Knowledge Discovery and Data Mining}}. \bibinfo{publisher}{Association for
  Computing Machinery}, \bibinfo{address}{New York, NY, USA},
  \bibinfo{pages}{1607--1615}.
\newblock


\bibitem[\protect\citeauthoryear{Shi, Chen, and Cartlidge}{Shi
  et~al\mbox{.}}{2021}]%
        {shi2021lobrm}
\bibfield{author}{\bibinfo{person}{Zijian Shi}, \bibinfo{person}{Yu Chen},
  {and} \bibinfo{person}{John Cartlidge}.} \bibinfo{year}{2021}\natexlab{}.
\newblock \showarticletitle{The {LOB} Recreation Model: Predicting the Limit
  Order Book from {TAQ} History Using an Ordinary Differential Equation
  Recurrent Neural Network}. In \bibinfo{booktitle}{\emph{Proceedings of the
  AAAI Conference on Artificial Intelligence}}, Vol.~\bibinfo{volume}{35(1)}.
  \bibinfo{publisher}{AAAI Press}, \bibinfo{address}{Palo Alto, CA},
  \bibinfo{pages}{548--556}.
\newblock


\bibitem[\protect\citeauthoryear{Sobol}{Sobol}{2001}]%
        {sobol2001global}
\bibfield{author}{\bibinfo{person}{Ilya~M Sobol}.}
  \bibinfo{year}{2001}\natexlab{}.
\newblock \showarticletitle{Global sensitivity indices for nonlinear
  mathematical models and their Monte Carlo estimates}.
\newblock \bibinfo{journal}{\emph{Mathematics and computers in simulation}}
  \bibinfo{volume}{55}, \bibinfo{number}{1-3} (\bibinfo{year}{2001}),
  \bibinfo{pages}{271--280}.
\newblock


\bibitem[\protect\citeauthoryear{Stanley, Plerou, and Gabaix}{Stanley
  et~al\mbox{.}}{2008}]%
        {stanley2008statistical}
\bibfield{author}{\bibinfo{person}{H~Eugene Stanley}, \bibinfo{person}{Vasiliki
  Plerou}, {and} \bibinfo{person}{Xavier Gabaix}.}
  \bibinfo{year}{2008}\natexlab{}.
\newblock \showarticletitle{A statistical physics view of financial
  fluctuations: Evidence for scaling and universality}.
\newblock \bibinfo{journal}{\emph{Physica A: Statistical Mechanics and its
  Applications}} \bibinfo{volume}{387}, \bibinfo{number}{15}
  (\bibinfo{year}{2008}), \bibinfo{pages}{3967--3981}.
\newblock


\bibitem[\protect\citeauthoryear{Vyetrenko, Byrd, Petosa, Mahfouz, Dervovic,
  Veloso, and Balch}{Vyetrenko et~al\mbox{.}}{2020}]%
        {vyetrenko2020get}
\bibfield{author}{\bibinfo{person}{Svitlana Vyetrenko}, \bibinfo{person}{David
  Byrd}, \bibinfo{person}{Nick Petosa}, \bibinfo{person}{Mahmoud Mahfouz},
  \bibinfo{person}{Danial Dervovic}, \bibinfo{person}{Manuela Veloso}, {and}
  \bibinfo{person}{Tucker Balch}.} \bibinfo{year}{2020}\natexlab{}.
\newblock \showarticletitle{Get real: Realism metrics for robust limit order
  book market simulations}. In \bibinfo{booktitle}{\emph{Proceedings of the
  First ACM International Conference on AI in Finance}}.
  \bibinfo{publisher}{ACM}, \bibinfo{address}{New York, NY},
  \bibinfo{pages}{1--8}.
\newblock


\bibitem[\protect\citeauthoryear{Wah and Wellman}{Wah and Wellman}{2016}]%
        {wah2016latency}
\bibfield{author}{\bibinfo{person}{Elaine Wah} {and} \bibinfo{person}{Michael~P
  Wellman}.} \bibinfo{year}{2016}\natexlab{}.
\newblock \showarticletitle{Latency arbitrage in fragmented markets: A
  strategic agent-based analysis}.
\newblock \bibinfo{journal}{\emph{Algorithmic Finance}} \bibinfo{volume}{5},
  \bibinfo{number}{3-4} (\bibinfo{year}{2016}), \bibinfo{pages}{69--93}.
\newblock


\bibitem[\protect\citeauthoryear{Wang, Hoang, Vorobeychik, and Wellman}{Wang
  et~al\mbox{.}}{2021}]%
        {wang2021spoofing}
\bibfield{author}{\bibinfo{person}{Xintong Wang}, \bibinfo{person}{Christopher
  Hoang}, \bibinfo{person}{Yevgeniy Vorobeychik}, {and}
  \bibinfo{person}{Michael~P Wellman}.} \bibinfo{year}{2021}\natexlab{}.
\newblock \showarticletitle{Spoofing the limit order book: A strategic
  agent-based analysis}.
\newblock \bibinfo{journal}{\emph{Games}} \bibinfo{volume}{12},
  \bibinfo{number}{2} (\bibinfo{year}{2021}), \bibinfo{pages}{46}.
\newblock


\bibitem[\protect\citeauthoryear{Yagemann, Chung, Uzun, Ragam, Saltaformaggio,
  and Lee}{Yagemann et~al\mbox{.}}{2020}]%
        {yagemann2020feasibility}
\bibfield{author}{\bibinfo{person}{Carter Yagemann}, \bibinfo{person}{Simon~P
  Chung}, \bibinfo{person}{Erkam Uzun}, \bibinfo{person}{Sai Ragam},
  \bibinfo{person}{Brendan Saltaformaggio}, {and} \bibinfo{person}{Wenke Lee}.}
  \bibinfo{year}{2020}\natexlab{}.
\newblock \showarticletitle{On the feasibility of automating stock market
  manipulation}. In \bibinfo{booktitle}{\emph{Annual Computer Security
  Applications Conference}}. \bibinfo{publisher}{Association for Computing
  Machinery}, \bibinfo{address}{New York, NY, USA}, \bibinfo{pages}{277--290}.
\newblock


\bibitem[\protect\citeauthoryear{Zhou and Lai}{Zhou and Lai}{2009}]%
        {zhou2009herding}
\bibfield{author}{\bibinfo{person}{Rhea~Tingyu Zhou} {and}
  \bibinfo{person}{Rose~Neng Lai}.} \bibinfo{year}{2009}\natexlab{}.
\newblock \showarticletitle{Herding and information based trading}.
\newblock \bibinfo{journal}{\emph{Journal of Empirical Finance}}
  \bibinfo{volume}{16}, \bibinfo{number}{3} (\bibinfo{year}{2009}),
  \bibinfo{pages}{388--393}.
\newblock


\end{thebibliography}

\end{document}